\documentclass[acmlarge, screen]{acmart}
\PassOptionsToPackage{table,xcdraw}{xcolor}
\usepackage{hyperref}

\AtBeginDocument{%
  \providecommand\BibTeX{{%
    \normalfont B\kern-0.5em{\scshape i\kern-0.25em b}\kern-0.8em\TeX}}}
    
\usepackage{enumitem}
\setlist[itemize]{leftmargin=*}
\usepackage{tabularx}
\usepackage{colortbl}
\usepackage{multirow,multicol}
\usepackage{booktabs}
\usepackage{multirow}
\usepackage{caption}
\usepackage{graphicx}
\usepackage{subcaption}
\usepackage{placeins}
\usepackage{float}
\usepackage{balance}
\usepackage{xspace}
\usepackage[table]{xcolor}
\usepackage{lscape}
\usepackage{siunitx}
\usepackage{soul}
\usepackage{hhline}

\definecolor{lightgray}{HTML}{EFEFEF}

\setcopyright{cc}
\setcctype{by}
\acmJournal{IMWUT}
\acmYear{2025} \acmVolume{9} \acmNumber{3} \acmArticle{92} \acmMonth{9} \acmPrice{}\acmDOI{10.1145/3749501}

\newcommand{\dataset}{\textit{SenseSeek}\xspace}
\newcommand{\task}[1]{{\small\sffamily #1}\xspace}

\newcommand{\eyeopen}{\task{EYEOPEN}}
\newcommand{\eyeclose}{\task{EYECLOSE}}
\newcommand{\ifn}{\task{IN}}
\newcommand{\qf}{\task{QF}}
\newcommand{\qs}{\task{QS}}
\newcommand{\qss}{\task{QS-S}}
\newcommand{\qst}{\task{QS-T}}
\newcommand{\rj}{\task{RJ}}
\newcommand{\rjr}{\task{RJ-R}}
\newcommand{\rjl}{\task{RJ-L}}

\usepackage{amsmath}
\DeclareMathOperator*{\argmax}{\arg\!\max}

\begin{document}

\title{\dataset Dataset: Multimodal Sensing to Study Information Seeking Behaviors
}

\author{Kaixin Ji}
\orcid{0000-0002-4679-4526}
\affiliation{%
  \institution{RMIT University} 
  \city{Melbourne}
  \country{Australia}
}
\email{kaixin.ji@student.rmit.edu.au}

\author{Danula Hettiachchi}
\orcid{0000-0003-3875-5727}
\affiliation{
\institution{RMIT University}
  \city{Melbourne}
  \country{Australia}
}

\email{danula.hettiachchi@rmit.edu.au}

\author{Falk Scholer}
\orcid{0000-0001-9094-0810}
\affiliation{
\institution{RMIT University}
  \city{Melbourne}
  \country{Australia}
}
\email{falk.scholer@rmit.edu.au}

\author{Flora D.~Salim}
\orcid{0000-0002-1237-1664}
\affiliation{%
  \institution{The University of New South Wales} 
  \city{Sydney}
  \country{Australia}
}
\email{flora.salim@unsw.edu.au}

\author{Damiano Spina}
\orcid{0000-0001-9913-433X}
\affiliation{
\institution{RMIT University}
  \city{Melbourne}
  \country{Australia}
}
\email{damiano.spina@rmit.edu.au}

\setlength{\intextsep}{10pt plus 2pt minus 2pt}

\begin{abstract}
Information processing tasks involve complex cognitive mechanisms that are shaped by various factors, including individual goals, prior experience, and system environments.
Understanding such behaviors requires a sophisticated and personalized data capture of how one interacts with modern information systems (e.g., web search engines). 
Passive sensors, such as wearables, capturing physiological and behavioral data, have the potential to provide solutions in this context.
This paper presents a novel dataset, \dataset, designed to evaluate the effectiveness of consumer-grade sensors in a complex information processing scenario: {searching via systems (e.g., search engines), one of the common strategies users employ for information seeking.}
The \dataset dataset comprises data collected from 20 participants, 235 trials of the stimulated search process, 940 phases of stages in the search process, including the realization of Information Need (\ifn), Query Formulation (\qf), Query Submission by Typing (\qst) or Speaking (\qss), and Relevance Judgment by Reading (\rjr) or Listening (\rjl).
The data includes Electrodermal Activities (EDA), Electroencephalogram (EEG), PUPIL, GAZE, and MOTION data, which were captured using consumer-grade sensors. It also contains 258 features extracted from the sensor data, the gaze-annotated screen recordings, and task responses. 
We validate the usefulness of the dataset by providing baseline {analyses} on the impacts of different \textit{cognitive intents} and \textit{interaction modalities} on the sensor data, and \textit{effectiveness} of the data in discriminating the search stages. 
To our knowledge, \dataset is the first dataset that {characterizes multiple stages} involved in information seeking with physiological signals collected from multiple sensors. We hope this dataset can serve as a reference for future research on information-seeking behaviors.

\end{abstract}
\settopmatter{printfolios=true}


\maketitle
\renewcommand{\shortauthors}{Kaixin Ji~et~al.}

\section{Introduction}

Information is ubiquitous~\cite{fidel2012human}, and modern information access systems such as web search engines or intelligent assistants have evolved to become increasingly diverse, offering greater mobility and multiple ways of interaction. In this context, wearable sensors that capture physiological and behavioral data have demonstrated remarkable capabilities in enhancing human-computer interaction and improving daily life applications. There is growing research interest in leveraging these wearable sensors to explore novel questions in information interaction, including the detection of user interest~\cite{arapakis2017interest, ashlee2016engaged} and engagement during video browsing~\cite{he2023understanding, banik2024emotions}, recognition of cognitive preference~\cite{bias2023boon, moravec2018fake} and emotional responses~\cite{Gebhardt2024socialmedia} on social media content, and understanding various information needs~\cite{Dominika2022information}.

Information only becomes meaningful when individuals extract and interpret it \cite{fidel2012human}. This interpretation process engages complex cognitive mechanisms, such as attention allocation, memory retrieval, reasoning, and decision-making \cite{moshfeghi2018search, ji2024characterizing}, shaped by various factors, including individual goals, prior experience, and system environments \cite{fidel2012human, nahl2007social}. Physiological data offers unique insights as it responds sensitively to different activities and variables, while physiological states themselves can also influence these activities. Given these intricate relationships between physiological data and contextual factors, careful examination of their interactions is essential \cite{riedl2014towards}. Moreover, wearable sensors are becoming smaller and more compact, as seen in new devices like earphone-style EEGs that have fewer sensors \cite{Zeydabadinezhad_2024, kaveh2024wireless}. Although these smaller devices are more convenient to wear, researchers need to balance this convenience with how well they can collect accurate and reliable data.

\paragraph*{Rationale for a new dataset}

In this paper, we focus on a complex information processing scenario: information seeking, {an iterative problem-solving process where users identify gaps in their knowledge and employ various strategies to address them \cite{kuhlthau2005information}. Searching via systems (e.g., search engines) is one of the common strategies users employ, involving multiple downstream tasks (referred to as search stages in this paper) that each carry different cognitive intents.}
A simplified explanation of this process is as follows, summarized from \cite{kuhlthau2005information, ji2024characterizing}. The user first recognizes a problem regarding their need for information and then reaches the system; this stage is termed \textit{the realization of Information Need (\ifn)}. Next, they think about the query, \textit{Query Formulation (\qf)}, and submit, \textit{Query Submission (\qs)}. When they receive the results from the system, they evaluate both the information contents and the quality of the search session; this stage is termed \textit{Relevance Judgment (\rj)}.
Given that these stages engage different cognitive processes, they may be reflected in bodily responses, which leads to our first research question, \textbf{RQ1: To what extent do different cognitive intents across search stages influence users' physiological responses?}
Moreover, the information systems have been diversified by providing different interaction modalities, i.e., Query Submission by Typing (\qst) or by Speaking (\qss) and Relevance Judgment by Reading (\rjr) or by Listening (\rjl).
As these modalities engage different sensory and motor systems, we propose our second research question, \textbf{RQ2: How do different interaction modalities influence users' physiological responses?}
On the other side, given the growing trend of employing physiological and behavioral data in information activities, especially in real-world settings (e.g., social media usage~\cite{Gebhardt2024socialmedia}, everyday workspace~\cite{kosmyna2019attentivuworkplace, mirjafari_differentiating_2019}), it is essential to understand the baseline performances of these measurements. This leads us to our third research question, \textbf{RQ3: How effectively can physiological and behavioral data detect and distinguish between specific search stages in information seeking processes?}

Therefore, we introduce the \dataset dataset, a comprehensive dataset comprising multimodal sensing data for studying information seeking behaviors. To the best of our knowledge, \dataset is the first dataset to analyze downstream searching tasks (i.e., search stages) using physiological and behavioral data collected using multiple consumer-grade sensors. It includes data from 20 participants who completed a controlled lab-based study simulating a mock-up information search scenario, divided into 6 search stages with 4 different interaction modalities. The user study content was carefully curated to ensure objectivity by minimizing the influence of subjective factors such as feelings, attitudes, and relevance.
The data comprises a range of physiological and behavioral data, including Electrodermal Activity (EDA), Electroencephalogram (EEG), MOTION (including wrist and head motion) from wearable devices, and PUPIL and GAZE from a screen-mounted eye tracker. It includes both raw and cleaned versions of the data, along with 258 extracted features. To provide contextual richness, it also contains the task materials, self-rated perceptions, and gaze-annotated screen recordings.

The contribution of this paper is 4-fold:
\begin{enumerate}
    \item Introducing the \dataset dataset, consisting of multiple sensor data (raw and cleaned) and extracted features, self-rated perceptions, gaze-annotated screen recordings from 20 participants performing mock-up information searching tasks;
    \item Understanding the physiological responses by the influence of cognitive intents and interaction modalities in a sequence of search stages by a highly controlled lab study;
    \item Assessing the effectiveness of physiological and behavioral data captured by consumer-grade sensors for monitoring information behavior;
    \item Proposing an experimental setup for data collection during information seeking can be replicated across diverse experimental settings, from naturalistic information search to conversational-style interaction, or multi-media information-access interaction.  
\end{enumerate}

\section{Related Work}
\label{sec:relatedwork}
This section introduces the background about information seeking and passive sensing, and discusses the necessity of establishing a baseline of the effectiveness of sensor data in this context. 

\subsection{Human Information Interaction \& Information Seeking}

Introduced by \citet{fidel2012human}, Human Information Interaction (HII) investigates how users interact with and make sense of information. HII is a broad and multidisciplinary field. Based on the research problem, HII has often been segmented and incorporated into specialized areas, such as Human-Computer Interaction (HCI), Information Retrieval (IR), or Information Behavior studies.

Information seeking is a sub-area in HII \cite{fidel2012human}. Information search is a process in which a user interacts with an information access system (e.g., a web search engine) to fulfill their need for information (i.e., information need) \cite{kuhlthau2005information}. 
In the field of IR, this process is usually described as an iterative process which consists of a sequence of search stages: the realization of Information Need (\ifn), Query Formulation (\qf), Query Submission (\qs) and Relevance Judgment (\rj) \cite{kuhlthau2005information}. 
Circumstances, goals, and motivations might drive user to seek particular information to address a gap in their knowledge or understanding (\ifn). 
In a lab study, a backstory describing the scenario is usually provided for this purpose.
They then plan about what query can help obtain the desired results from the system (\qf) and proceed to execute the plan (\qs). Finally, they evaluate whether the results meet their information goals (\rj).

To enhance user experience and accurately meet user needs in various contexts, wearable sensors might have great potential to offer a portable and accessible solution \cite{lopatovska2011theories}.

\subsection{Understanding Information Behaviors with Passive Sensing}

Passive sensing, including the use of behavioral and physiological sensors, offers a unique way to study information behaviors by capturing real-time, naturalistic data. This approach provides deeper insights into how individuals engage with and process information in everyday contexts.

Eye-tracking data, such as eye movements, is commonly used in information behavior studies to visualize attention and assess user judgments on relevance~\cite{buscher2012attentive}, task goals~\cite{cole2014task}, and interests~\cite{jacob2018gaze}. Additionally, motion data, including wrist and head movements, is valuable for understanding emotional states and engagement. For instance, wrist movements have provided insights into learning engagement~\cite{ngage2020gao}. In a study about social media use, \citet{Gebhardt2024socialmedia} found that behavioral features have outperformed the physiological features in emotion detection.

Physiological data, including skin conductance, heart rate, pupil dilation, and brain activity, have been widely used to explore cognitive or affective states in information behavior studies. For example, 
\citet{gupta2022recall} used physiological data and eye activity for recognition of emotional memory recalls.  
\citet{jimenez2018using} measured workload changes during web browsing with multiple physiological data. Their classification models using EEG data outperformed those excluding it. \citet{moshfeghi2019towards} have utilized brain activity to detect information need realization, while \citet{ye2022towards} explored EEG responses when individuals identify keywords linked to information needs. \citet{moshfeghi2013effective} found that combining physiological data with task dwell time significantly improved the discrimination of search intentions.
\citet{wu2017predicting} demonstrated that EDA could predict online shopping satisfaction, while \citet{White2017881} showed that heart rate features were closely linked to user interest on search pages. \citet{ashlee2016engaged} found that skin conductance was associated with interest, while heart rate indicated frustration during search tasks.

\subsection{Dynamics of Information Interaction \& Sensor Data}
Information activity centers on the interaction between users and information \cite{fidel2012human}.
Beyond the inherent characteristics of the information contents, various factors during user interactions with information access systems might also influence sensor data. These influences, which may affect the generalizability of results, remain insufficiently understudied.

First of all, the \emph{cognitive intent}.
As previously discussed, during information searching, users progress through the information stages with varying cognitive intents. Research found that these shifts in intent trigger diverse affective and cognitive processes \cite{moshfeghi2016understanding}, which may, in turn, influence physiological responses.

Secondly, the \emph{interaction modalities}.
To accommodate evolving user demands, the information access systems have been renovated into multi-media platforms offering diverse interaction methods. These include not only traditional screen-based interactions but also voice-based interactions and their combinations.
Different interaction methods undergo distinct internal processes in humans, evoking varied psychological responses. 
For example, the single-channel EEG data~\cite{so2017evaluation} or physiological responses~\cite{kshitij2020cognitive} are used to assess mental workloads across contexts with varying cognitive and motor demands. \citet{iadarola2021analysis} found that EDA captured by the Empatica E4 wristband could differentiate the emotions in pleasant and unpleasant acoustic stimuli.
It is yet understudied whether these responses impact the physiological data captured by consumer-level wearable sensors.
In this paper, we focus on comparing different interaction modalities in submitting or receiving information. Specifically, submitting the query to the system by typing (\qst) or by speaking (\qss), and receiving the search results from the system in text format for reading (\rjr) or audio format for listening (\rjl).

Furthermore, physiological and behavioral data have their unique advantages and limitations. Firstly, since different information stages require various physical actions, behavioral data may exhibit more observable differences among search stages compared to physiological data. However, physiological data can provide insights into users' internal states such as stress levels, cognitive load, and emotional responses, which behavioral data alone cannot capture.
Secondly, physiological data have varying response times to stimuli~{\cite{schmidt_wearable-based_2019}}. This variability is particularly relevant in the context of information searching, where the duration of search stages can range from a few seconds to over a minute, unlike tasks in other studies with consistent engagement durations, e.g., \cite{Gebhardt2024socialmedia}. 
There is a growing interest in incorporating these sensing data into advanced analytical applications (e.g., building predictive models), but what is less clear is how these variants, instead of the experimental stimulus, impact model performance.

\subsection{Multimodal Sensing Datasets}

A body of recent research has established valuable datasets for studying physiological and wearable sensor data across diverse contexts. The DEAP dataset~\cite{koelstra2011deap} laid the groundwork for emotion analysis using physiological signals, while newer contributions have expanded into specialized domains. The Emognition dataset~\citet{saganowski2022emognition} and \citet{banik2024emotions} focused on continuous emotion annotation during video consumption, and EEG-SVRec~\cite{zhang2024eegsv} for short-video recommendation. \citet{bota2024real} focused on the group emotion experiences during movie sessions in a more naturalistic setting.
VREED~\cite{tabbaa2022vr} incorporated virtual reality contexts with eye tracking, and LAUREATE~\cite{LAUREATE2023Laporte} targeted memory augmentation applications. WEAR~\cite{bock2024wear} focused on outdoor sports activity recognition. \citet{zhang2024zone} worked on collaborative cognitive experience using EEG. These datasets demonstrate the evolution toward multimodal sensing approaches in naturalistic settings, with a focus on solving practical problems. It also highlights a persistent challenge in the need for standardized annotation.

\section{Data Collection Methodology}
This section introduces the details of the data collection procedure.

\subsection{Participants}
We collected the data from 20 participants (12 male, 8 female).
A summary of their demographic attributes is given in Table~\ref{tab:demogra}. 
All participants were required to have at least professional working proficiency in English to ensure a minimum of additional effort was involved when completing the study, which was presented in English.

The study received ethics approval from {RMIT University}, and the participants provided written consent prior to the experiment. A total of 29 participants were recruited. After the experiment, the participants were contacted again for follow-up consent to release the raw data, and 20 out of 29 participants provided consent.

\begin{table}[hp!]
\caption{{Demographic of participants in the dataset.}}
\label{tab:demogra}
\resizebox{0.8\textwidth}{!}{
\begin{tabular}{lclc|lc|lc|lc}
\toprule
\multicolumn{2}{c|}{\textbf{Gender}}           & \multicolumn{2}{c|}{\textbf{Age}} & \multicolumn{2}{c}{\textbf{English Proficiency}} & \multicolumn{2}{|c}{\textbf{Right-Handed}} & \multicolumn{2}{|c}{\textbf{Wear-Glasses}}\\ \midrule
Male                 & \multicolumn{1}{c|}{12} & 18--24 years old      & 5& Professional                   & 4 & Y&18 & Y&12\\
Female               & \multicolumn{1}{c|}{8} & 25--34 years old      & 14& Full Professional              & 11 & N&2& N&8\\
\multicolumn{1}{c}{} & \multicolumn{1}{c|}{}   & 35--44 years old      & 1       & Native English                 & 5 & & & &\\ \midrule
\multicolumn{4}{l}{\textbf{Total Number of participants}}                         & \multicolumn{6}{l}{20}\\ \bottomrule
\end{tabular}}
\end{table}

\begin{figure}[htbp!]
    \centering
    \includegraphics[width=\linewidth]{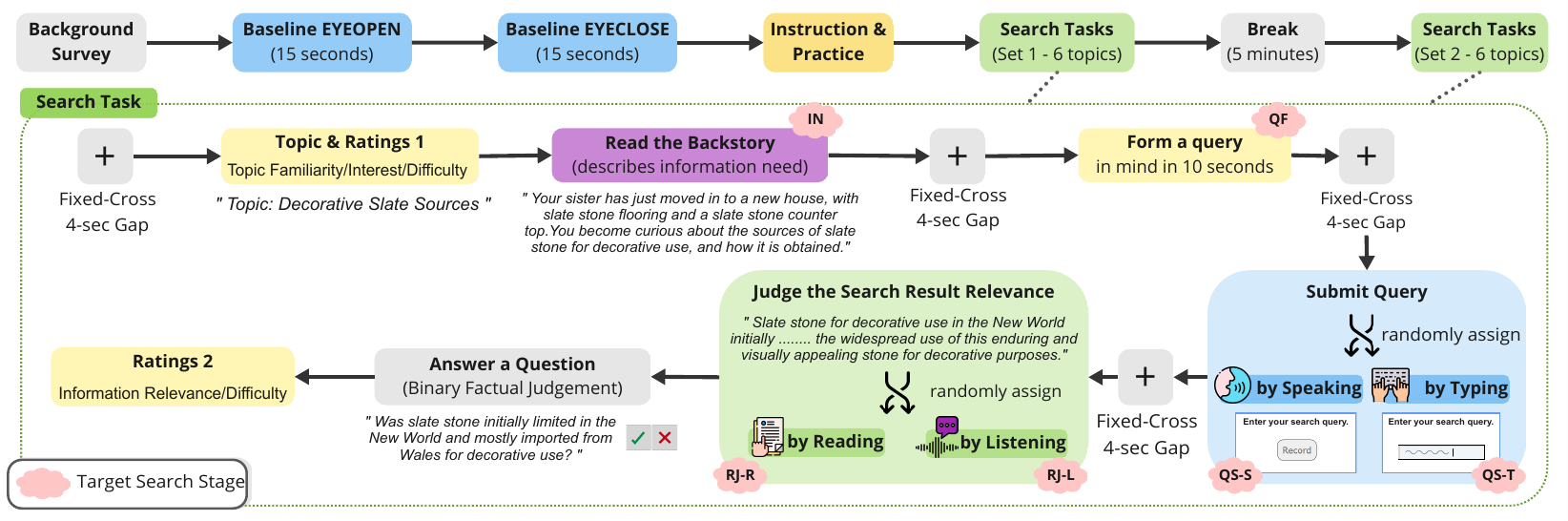} 
    \caption{User study procedure. The top section outlines the overall procedure, while the bottom details the search task procedure. Topic order and interaction modalities are randomized. Example contents for one topic are shown in \textit{italic} font. The target stages are labeled with their abbreviations: Information Need (\ifn), Query Formulation (\qf), Query Submission by Typing (\qst) or Speaking (\qss), and Relevance Judgment by Reading (\rjr) or Listening (\rjl).}
    \Description{A flowchart illustrating the sequence of tasks participants completed during the study. It includes baseline recordings, topic presentation, query formulation, submission (typing or speaking), relevance judgment (reading or listening), and ratings. The diagram highlights randomized topic order and interaction modalities.}
    \label{fig:exp1fig}
\end{figure}
\raggedbottom

\subsection{Material}
The topics and the corresponding backstories (i.e., task scenarios) are selected from the TREC2002-InformationNeed dataset~\cite{informationneed2014}. This dataset contains backstories corresponding to different topics from the TREC2002-4 topic set, and categorizes them into three labels in terms of cognitive complexity.
We use the topics from the Understanding category, which requires the participants to find information and gain some understanding of the topics. 
After removing the topics related to crises, wars, conspiracy, or politics, which might trigger subjective feelings, we select 12 topics and their corresponding backstories.

We reviewed and adjusted the backstories and the search results based on Flesch-Kincaid readability scores and word counts. All search results were converted into an audio format with the Google text-to-speech API for the listening tasks. 
As a result, the backstories have an average of 40$\pm$1 words, the search results have an average 148$\pm$3 word count with a readability level of 12, and the corresponding audios average 64$\pm$4 seconds.

\subsection{Search Task}

The search task is inspired and modified based on the experiment in \citet{moshfeghi2018search}, which contains the following search stages: Information Need (\ifn), Query Formulation (\qf), Query Submission (\qs), and Relevance Judgment (\rj). Note that as we are interested in the different responses among search stages rather than particular evaluation criteria, only one pre-defined highly-relevant search result is provided at \rj for each topic.

Each task starts with a 4-second blank, where the participants are asked to look at the fixed cross in the middle of the screen. A topic is shown afterward. And the participants need to rate their interests, familiarity, and perceived difficulty regarding the topic, with a 5-point Likert scale 
(as listed in Appendix~\ref{apx:add_material}). 
After that, a backstory that evokes their information need is presented. The participants are then given 10 seconds to form a search query in mind (\qf). Next, they follow an instruction to submit the query by speaking (\qss) or typing (\qst). Then the participants receive the search result by either a text paragraph to read (\rjr) or an audio to listen (\rjl). In the end, the participants need to answer a binary judgment question to ensure engagement and rate perceived relevance and difficulty in understanding the search result, with a 5-point Likert scale (as listed in Appendix~\ref{apx:add_material}). Apart from \qf, there was no time constraint on completing the task.

To avoid the physiological responses overlapping between the stages, the 4-second gap is also provided before the three interested stages, i.e., \qf, \qs, \rj. 
The sequences of topics and the combination of interaction modalities are randomized.

\subsection{Procedure}

After calibrating all sensors, the participants need to answer a background survey. The survey asked questions that might impact the data quality, e.g., handiness, caffeine intake, sleep hours, and spectacular. Next, the participants are instructed to have a 15-second eyes-open (\eyeopen) and 15-second eyes-close (\eyeclose) section to collect the baseline, followed by a training section containing the instruction and two practice tasks. Then the participants proceed to the main search tasks (12 tasks in total). There is a 5-minute break provided in the middle. After completing the whole experiment, participants are asked some questions regarding their experience during the experiment to reveal some undetected activities. For example, `do you feel the 10 seconds given for query formulation is unnecessary or useful?'. 

The details in the search tasks, e.g, the time given to query formulation and the total number of tasks, were pilot-tested with 4 additional participants.

\subsection{Apparatus and Setup}

\begin{figure}[htbp!]
    \centering
    \begin{subfigure}[t]{.45\linewidth}
        \centering
        \includegraphics[width=\linewidth, trim={0 5cm 0 0}, clip]{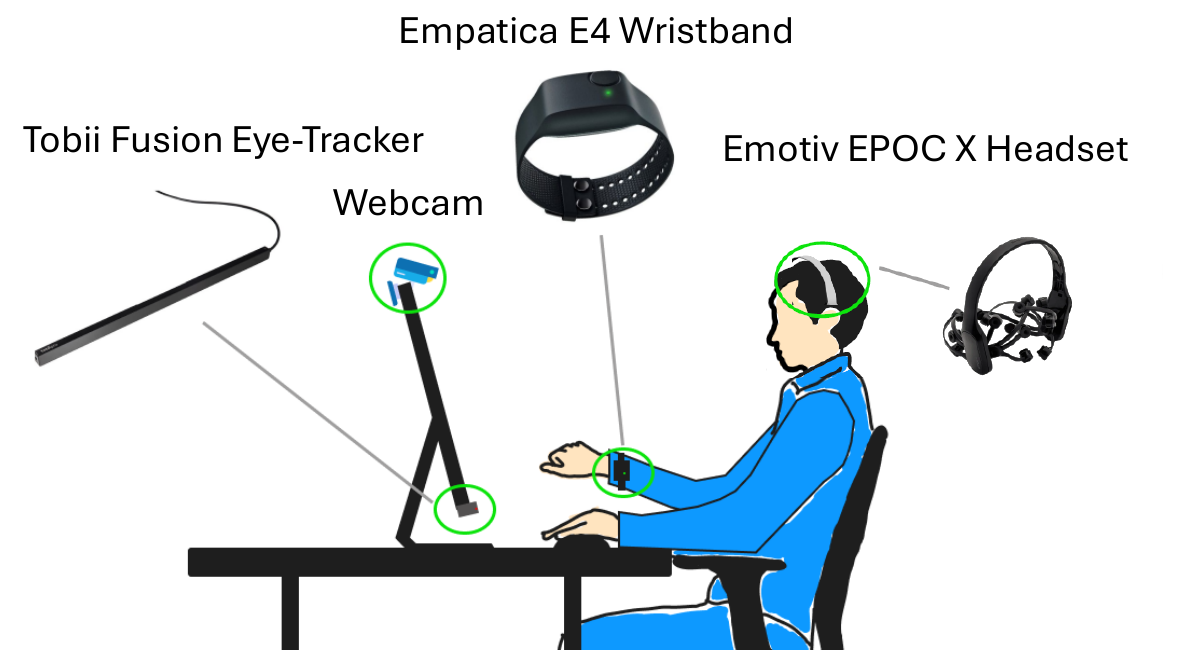}
        \caption{{Experimental setup.}}
    \end{subfigure}
    ~
    \begin{subfigure}[t]{.5\linewidth}
        \centering
        \includegraphics[width=0.4\linewidth]{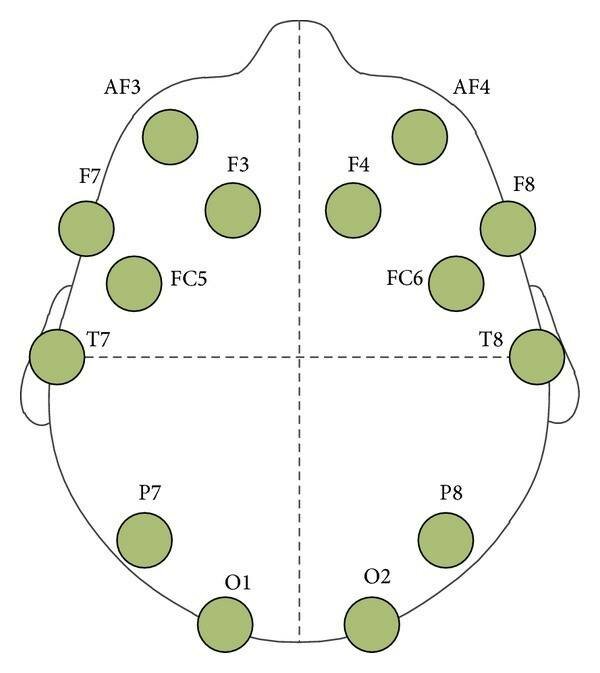}
        \caption{The EEG electrode locations (adapted from \cite{gupta2019enhancing}).}
    \end{subfigure}
    \vspace{-8pt}
        \caption{Experimental setup and apparatus.}
        \Description{The left figure illustrates a person sitting in front of a computer and wearing various devices for data collection. The right figure depicts EEG electrode placements on a head.}
        \label{fig:setup}
\end{figure}

The list of apparatus and collected data is presented in Table~\ref{tab:data}, and the setup is presented in Figure~\ref{fig:setup}.
The study is conducted in an illuminated room, which includes a desktop PC mounted with an eye-tracker and a web camera. All participants are provided with a standard right-handed computer mouse to complete the tasks. The participant sits in front of the computer and wears the wristband on the left hand. The instructor cleans the electrodes and the participant's skin on the inner and outer wrist with alcohol wipes \cite{eda2021baba}, and helps the participants to wear the headset and adjusts the position of the electrodes. The experiment is created on the Qualtrics\footnote{\url{https://www.qualtrics.com/about/}} platform.

\begin{table}[ht!]
\caption{List of sensor data collected in the dataset.}
\label{tab:data}
\resizebox{\textwidth}{!}{
    \centering
    \begin{tabular}{@{}llllll@{}}
    \toprule
    \rowcolor[HTML]{EFEFEF} 
    &&\textbf{Sampling}&&&\\
    \rowcolor[HTML]{EFEFEF} 
    \textbf{Description}         & \textbf{Data} & \textbf{Rate}& \textbf{Channels}                                               & \textbf{Comments}                      & \textbf{Apparatus}                        \\ \midrule
    
    Electrodermal Activity       & EDA           & 4Hz                    & 1                                                               & Unit in microsiemens (uS)              & \multirow{2}{*}{Empatica E4 Wristband\footnote{\url{https://www.empatica.com/en-int/research/e4/}Note that the E4 wristband also has a PPG sensor to collect heart rate data. However, this data was not collected in this study due to an error in wearing the device.}}              \\
    Wrist Motion - accelerometer & MOTION       & 32Hz                   & 3 ($x,y,z$)                                                       & Data in {[}-2g, 2g{]} range      &                                           \\
    \midrule
    Head Motion - accelerometer  & \multirow{3}{*}{MOTION}       & 128Hz                   & 3 ($x,y,z$)                                                       & Data in {[}-16g, 16g{]} range    & \multirow{4}{*}{Emotive EPOC X Headset\footnote{\url{https://www.emotiv.com/epoc-x/}}}           \\
    \multicolumn{1}{r}{- magnetometer}   &        & 128Hz                   & 3 ($x,y,z$)                                                       & Data in {[}-2000dps, 2000dps{]} range    &                                           \\
    \multicolumn{1}{r}{- quaternions}    &        & 128Hz                   & 1                                                               & Data in {[}-1g, 1g{]} range      &                                           \\
    Electroencephalogram         & EEG           & 128Hz                   & 14 (AF3, F7, F3, FC5,   & Unit in microvolt (uV)                 &                                           \\
    & & & T7, P7, O1, O2, P8, & & \\
    & & & T8, FC6, F4, F8, AF4) & & \\
    \midrule
    Gaze Movement                & GAZE          & 60Hz                   & 2 ($x,y$) x 2 (left, right)                                       & The coordinate of gaze point in screen & \multirow{2}{*}{Tobii Fusion Eye-tracker\footnote{\url{https://www.tobii.com/products/eye-trackers/screen-based/tobii-pro-fusion}}} \\
    Pupil Diameter               & PUPIL         & 60Hz                   & 1 x 2 (left, right)                                             & Unit in millimeters (mm)               &                                           \\ \bottomrule
    \end{tabular}
    }
\end{table}

\section{\dataset Dataset Description}
This section describes the dataset and the data cleaning and processing steps for the validation.

\subsection{Overview}

This dataset comprises data collected from 20 participants, each completing 12 search tasks, for a total of 240 trials. We excluded 5 trials due to unexpected environmental disturbances and equipment failures, leaving 235 trials in the dataset. Additionally, eye-tracking data for 2 participants were not recorded because of device errors.
The summary of the dataset is presented in Table~\ref{tab:dataset_size_full}.

\begin{table}[tb]
\caption{Table of Glossaries and Abbreviations. *Note that these are specialized terminologies used in the field of information behavior studies \cite{kelly2009methods}.}
\vspace{-8pt}
\label{tab:glossary}
\resizebox{\textwidth}{!}{
\begin{tabular}{@{}lll@{}}
\toprule
\rowcolor[HTML]{EFEFEF} 
\textbf{Stage Name}                       & \textbf{Abbreviation} & \textbf{Description} \\ \midrule
Baseline EYEOPEN                    & \eyeopen               &     A baseline task to relax and look at the screen with eyes open for 15 seconds.                 \\
Baseline EYECLOSE                   & \eyeclose              &      A baseline task to relax and close the eyes for 15 seconds.                \\
The realization of Information Need* & \ifn                   & \begin{tabular}[c]{@{}l@{}}Read the given backstory* (i.e., scenario) to understand the context and the need \\for information.\end{tabular}                                \\ 
Query Formulation*                   & \qf          &        Think about a search query for getting the desired information from the system.              \\ 
Query Submission* by Typing                         & \qst                  &    Type the search query and submit to the system.                \\
Query Submission* by Speaking                       & \qss                 &      Speak the search query and submit to the system.               \\
Relevance Judgment* by Reading                        & \rjr                 &    \begin{tabular}[c]{@{}l@{}}Receive a (highly relevant in this study) search result in text format to read,  and \\judge the relevance to the information need.\end{tabular} \\
Relevance Judgment* by Listening                      & \rjl                 &     \begin{tabular}[c]{@{}l@{}}Receive a search result in audio format to listen to, and judge the relevance to \\the information need.\end{tabular}                           \\ 
\bottomrule
\end{tabular}
}
\end{table}

\begin{table}[tb]
\caption{Summary of the \dataset dataset.}
\vspace{-8pt}
\label{tab:dataset_size_full}
\resizebox{\textwidth}{!}{
\begin{tabular}{@{}ccccccccccc@{}}
\toprule
\rowcolor[HTML]{EFEFEF} 
\multicolumn{11}{c}{\textbf{Recorded Contextual Data}}     \\ \midrule
\multicolumn{3}{l}{\textbf{Number of participants}}                                                                                                                     & \multicolumn{3}{l}{20}                                                          & \multicolumn{4}{l}{\textbf{Number of Topics}}                     & \multicolumn{1}{l}{12}                       \\
\multicolumn{3}{l}{\textbf{Number of Gaze-annotated Screen Recordings}}                                                                                                           & \multicolumn{3}{l}{20}                                                          & \multicolumn{4}{l}{\textbf{Number of Task Responses}}             & \multicolumn{1}{l}{235}                      \\ 
\toprule
\rowcolor[HTML]{EFEFEF} 
\multicolumn{11}{c}{\textbf{Recorded Sensing Data}}                                                                                                                                                                                                                                                                                                                          \\ \midrule
\multicolumn{2}{l}{\multirow{2}{*}{\textbf{Search Stages}}}                                                                                        & \multicolumn{2}{c}{\textbf{Baseline}} & \multirow{2}{*}{\textbf{IN}} & \multirow{2}{*}{\textbf{QF}} & \multicolumn{2}{c}{\textbf{QS}} & \multicolumn{2}{c}{\textbf{RJ}} & \multirow{2}{*}{\begin{tabular}[c]{@{}c@{}}\textbf{Total}\\(+ baseline)\end{tabular}} \\  \cmidrule(lr){3-4} \cmidrule(lr){7-8} \cmidrule(lr){9-10}
\multicolumn{2}{c}{}                                                                                                                               & \textbf{EYEOPEN}  & \textbf{EYECLOSE} &                              &                              & \textbf{QS-T}  & \textbf{QS-S}  & \textbf{RJ-R}  & \textbf{RJ-L}  &                                              \\ \midrule
\multicolumn{2}{l}{\textbf{Duration in seconds (sd)}}                                                                                              & 15.0 (0)          & 15.0 (0)          & 23.4 (14.1)                  & 10.0 (0)                     & 17.7 (10.1)    & 6.8 (2.5)      & 49.2 (19.8)    & 66.0 (4.4)     & -                                            \\ \midrule
\multirow{6}{*}{\textbf{\begin{tabular}[l]{@{}l@{}}Amount\\ of Data \\ Instance\end{tabular}}} & \multicolumn{1}{l}{\textbf{EEG}}          & 20                & 20                & 235                          & 235                          & 119            & 116            & 118            & 117            & \textbf{940 (+40)}                           \\ 
                                                                                               & \multicolumn{1}{l}{\textbf{MOTION} (head)} & 20                & 20                & 235                          & 235                          & 119            & 116            & 118            & 117            & \textbf{940 (+40)}                           \\ 
                                                                                               & \multicolumn{1}{l}{\textbf{MOTION} (wrist)} & 20                & 20                & 235                          & 235                          & 119            & 116            & 118            & 117            & \textbf{940 (+40)}                           \\  
                                                                                               & \multicolumn{1}{l}{\textbf{EDA}}            & 20                & 20                & 235                          & 235                          & 119            & 116            & 118            & 117            & \textbf{940 (+40)}                           \\ 
                                                                                               & \multicolumn{1}{l}{\textbf{GAZE} }          & 18                & NA                & 211                          & 211                          & 107            & 101            & 106            & 105            & \textbf{859 (+18)}                           \\ 
                                                                                               & \multicolumn{1}{l}{\textbf{PUPIL}}         & 18                & NA                & 211                          & 211                          & 107            & 101            & 106            & 105            & \textbf{859 (+18)}                           \\ \bottomrule
\end{tabular}}
\end{table}

\subsection{Sensor Synchronization}
The event timestamps of the task behaviors were recorded with the JavaScript library on Qualtrics. 
To synchronize all sensors, we convert the timestamps into ISO 8601 time format with milliseconds. The signals were aligned and segmented according to the event timestamps recorded.
An example of apparatus timeline and synchronization is presented in Figure~\ref{fig:exptimeline}.

\begin{figure}[htbp!]
    \centering
    \includegraphics[width=0.99\linewidth]{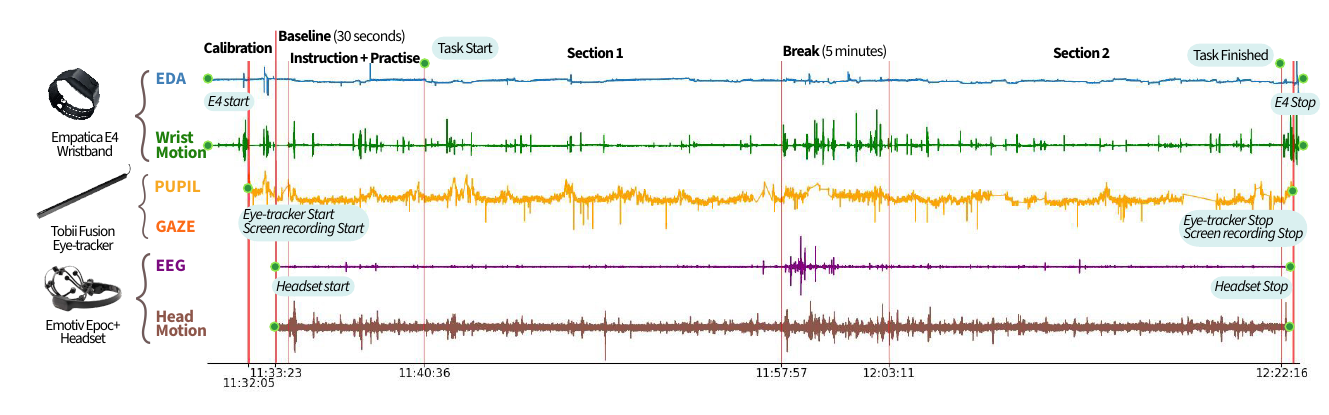}
    \caption{Example of Apparatus Timeline and Synchronization in the Data Collection.}
    \Description{A timeline diagram showing the synchronization of sensor data collection during different phases of the experiment, including calibration, baseline, task execution, and breaks.}
    \label{fig:exptimeline}
\end{figure}
\raggedbottom

\subsection{Data Pre-Processing}

All data was processed with the Python programming language. EEG {data were} processed using the MNE library \footnote{\url{https://mne.tools/stable/i}}. EDA and wrist motion data obtained from the wristband were processed using the NeuroKit2~\cite{makowski2021neurokit} library. 
PUPIL data was cleaned with the \textit{PyPlr}~\cite{Martin2021} library. 
{Data collected during the break section were} excluded from pre-processing. 

\paragraph*{EEG}

Following similar procedures to \citet{martinez2023impact}, \citet{ye2023relevance}, \citet{Gwizdka2017temporal}, EEG data were first re-referenced with the common average, denoised with a Butterworth filter (1 -- 40Hz), and removed the signal mean for the middle 90\% percentile.  
Next, to remove the artifacts (e.g., blinking), the data was further cleaned and interpolated with the Autoreject~\cite{jas2017autoreject} package and Independent Component Analysis (ICA) combined with ICLabel \cite{Li2022}.
The power spectral density (PSD) of each EEG channel was then calculated using Welch's method and hamming window and normalized \cite{kosonogov2023eeg, so2017evaluation, lee2020frontal}, for 4 frequency bands, $Theta$ (4--8Hz), $Alpha$ (8--13 Hz), $Beta$ (13--30Hz) and $LowGamma$ (25--40Hz) \cite{Gwizdka2017temporal}.

\paragraph*{EDA}
Following a similar procedure as by \citet{ngage2020gao, Gebhardt2024socialmedia}, 
EDA was first cleaned with a rolling median filter with a 3-second window (12 data points)~\cite{eda2021baba}, and then standardized using the $z$-score. Next, the convex optimization \textit{cvxEDA} method \cite{greco2016cvxeda} was applied to decompose the tonic value, i.e., the Skin Conductance Level (SCL), and the phasic value, i.e., the Skin Conductance Response (SCR).

\paragraph*{MOTION}
No pre-processing was done on the head motion data. {For the wrist motion, we }followed the procedure of \citet{ngage2020gao}. We first calculated the
magnitude from the 3 coordinates ($x,y,z$),
$|a|=\sqrt{x^2+y^2+z^2}$, then cleaned with a rolling median filter with 0.2-second window (6 data points).

\paragraph*{GAZE \& PUPIL}
The invalid gaze samples identified by the eye-tracker were removed. 
We used Tukey's method on the first derivatives to remove the noise caused by blinks from the left and right pupils, respectively. The samples with first derivatives lower than or higher than 3 standard deviations are removed.
Then, the cleaned data of both sides were combined by taking the arithmetic mean, and linear interpolation was applied to fill in the blink gaps. Finally, a zero-phase Butterworth filter (4Hz, $3^{rd}$) was applied to remove outliers \cite{Martin2021}.
We also calculated Relative Pupil Dilation (RPD), the relative changes of current pupil diameter compared to a baseline value (\eyeopen), following the formula by \citet{Gwizdka2017temporal}: $RPD_{t}^{i} = (P_{t}-P_{baseline}^{i})/{P_{baseline}^{i}}$ where $t$ is time, $i$ is participant, and baseline is the average pupil diameter during \eyeopen.

\subsection{Features Extraction}
We extracted a total of 258 features from both \emph{physiological} (EDA, EEG, and PUPIL) and \emph{behavioral} (MOTION and GAZE) data, following \cite{ngage2020gao, barral2015exploring, bias2023boon, Gwizdka2017temporal, bird2018study, Duchowski2020lowhigh}. The features are listed in Table~\ref{tab:features}. More details on the extraction are described in the Appendix~\ref{apx_sub:features}.

\paragraph*{Normalization}
Same as in \citet{Gebhardt2024socialmedia}, the normalization was done by per participant. 
The physiological features were normalized by subtracting the value in the baseline period (i.e., \eyeopen).
The behavioral features were normalized by subtracting the mean values overall for the whole experiment and then dividing by the standard deviation.

\begin{table}[tp!]
\caption{List of total 258 extracted features. Skin Conductance Level (SCL), Skin Conductance Response (SCR), standard deviation (sd). Relative Pupil Dilation (RPD). *Four EEG frequency band: $Theta$ (4--8 Hz), $Alpha$ (8--13 Hz), $Beta$(13--25 Hz) and $LowGamma$ (25--40 Hz) bands \cite{Gwizdka2017temporal}. **Four channel groups: frontal-left (AF3, F7, F3, FC5), frontal-right (FC6, F4, F8, AF4), back-left (P7, O1), and back-right (P8, O2) \cite{Gwizdka2017temporal}. ***statistics (5): five statistical description features include mean, max, min, sd, range. 
}
\label{tab:features}

\resizebox{\linewidth}{!}{
\begin{tabular}{@{}cclp{8cm}l@{}}
\toprule
\rowcolor[HTML]{EFEFEF} 
\textbf{Type}                  & \textbf{Group}                           & \textbf{Category}             & \textbf{Features}                                                                                              & \textbf{N. Features} \\ \midrule
\multirow{11}{*}{\rotatebox[origin=c]{90}{Physiological}} & \multirow{3}{*}{\textbf{EDA(31)}}        & \multirow{2}{*}{EDA}                          & statistics (5), median, variance, skewness, kurtosis                                                           & 9                    \\ 
  &     &  & statistics (5) of 1st/2nd derivative of EDA                                                                                                   & 10 (5x2)             \\ 
\cmidrule(l){3-5} 
                               &                                          & SCL                           & statistics (5), median, mean amplitude                                                                         & 7                    \\ \cmidrule(l){3-5} 
                               &                                          & SCR                           & statistics (5)                                                                                                 & 5                    \\ \cmidrule(l){2-5} 
                               & \multirow{3}{*}{\textbf{EEG(180)}}       & 14 channels                   & mean, sd, skewness, kurtosis, curve length, zero crossings, number of peaks, wavelet entropy                   & 112 (8x14)           \\ \cmidrule(l){3-5} 
                               &                                          & 4 frequency bands*                & the logarithms of the spectral power                                                                                               & 56 (4x14)            \\ \cmidrule(l){3-5} 
                               &                                          & 4 bands of 4 channel groups** & Inter-/Intra-hemispheric asymmetry power ratios between/within right-left groups                               & 12 (4+8)             \\ \cmidrule(l){2-5} 
                  &  \multirow{4}{*}{\textbf{PUPIL(25)}}               & \multirow{2}{*}{Pupil Diameter}                & statistics (5), median                                                                                         & 6                    \\  
                                             
                               &                                          &                      & LHIPA index~\cite{Duchowski2020lowhigh}                                                                                                          & 1                    \\ 
                                   \cmidrule(l){3-5} 
                               &                                          & \multirow{2}{*}{RPD}                           & statistics (5), 5 quarterlies (10, 25, 50, 75, 90)                 & 10                   \\
                               &               &       & statistics (5), 3 quarterlies (25, 50, 75) of 1st derivative of RPD                                                                    & 8                    \\ 
\midrule
\multirow{4}{*}{\rotatebox[origin=c]{90}{Behavioral}}   &  \multirow{3}{*}{\textbf{MOTION(13)}}            &   Wrist Motion  & statistics (5) of magnitude of acceleration                                                                                                 & 5                    \\ \cmidrule(l){3-5} 
                               & &  Head Motion  & mean of \textit{x, y, z} coordinates  of  angular speed/acceleration                                                        & 6 (3x2)              \\  
                               &                                          &    & mean, energy of  magnitude of acceleration                                                                                                   & 2                    \\ \cmidrule(l){2-5} 
                               & \multirow{1}{*}{\textbf{GAZE(6)}}             & fixation/saccade              & number of occurrence, total duration (normalized), mean duration of each occurrence & 6 (3x2)              \\                 \bottomrule
\end{tabular}
}
\end{table}

\section{Validation}
\label{sec:exploratory}

This paper aims to validate the \dataset dataset by understanding how different \textit{cognitive intents} \textbf{(RQ1)} and \textit{interaction modalities} \textbf{(RQ2)} across search stages influence physiological responses, and investigating \textit{the effectiveness of using physiological and behavioral data to distinguish search stages} \textbf{(RQ3)}. This section presents our validation methodology and findings organized according to these research questions.

Before conducting any analysis, we first {reviewed} the self-reported task perceptions to ensure the experimental conditions {met} our anticipation (Section~\ref{subsec:manipulation_check}).
To answer RQ1 and RQ2, we conducted the exploratory analysis with statistical tests in a within-subject setting. 
We first conducted non-parametric Friedman tests to compare across \eyeopen baseline, and all search stages (\ifn, \qf, \qst, \qss, \rjr, \rjl) for overall impacts (Section~\ref{subsec:responses_stages}). 

We then performed post-hoc analysis to compare stages with similar characteristics: \eyeopen, \ifn, and \rjr (different cognitive intents, same interaction modalities, RQ1, Section~\ref{subsec:influence_intents}), and \qst vs. \qss, and \rjr vs. \rjl (same cognitive intents, different interaction modalities, RQ2, Section~\ref{subsec:influence_modalities}).
The multiple comparisons were accounted for with Bonferroni correction. 
For RQ3, we conducted a classification analysis by developing machine learning models using features from individual and multiple sensor groups, and comparing the performances in discriminating all search stages.

\subsection{Task Completeness \& Manipulation Check}
\label{subsec:manipulation_check}

\paragraph*{Task Perceptions}

As the confounding variables, for instance, perceived difficulty~\cite{martinez2023impact}, interest~\cite{bias2023boon, arapakis2017interest}, or search result relevance~\cite{ye2023relevance}, could impact the physiological responses, the stimulus used in this study has been taken care to minimize the impact from these. In particular, 
we anticipate that i) all the topics should be easy to understand ($\leq 3.0$) and not familiar ($\leq 3.0$), ii) all the search results (in \rj) provided are easy to understand ($\leq 3.0$), and relatively relevant ($\geq 3.0$) to their submitted query.

\begin{figure}[htbp!]
    \centering
    \includegraphics[width=0.41\textwidth]{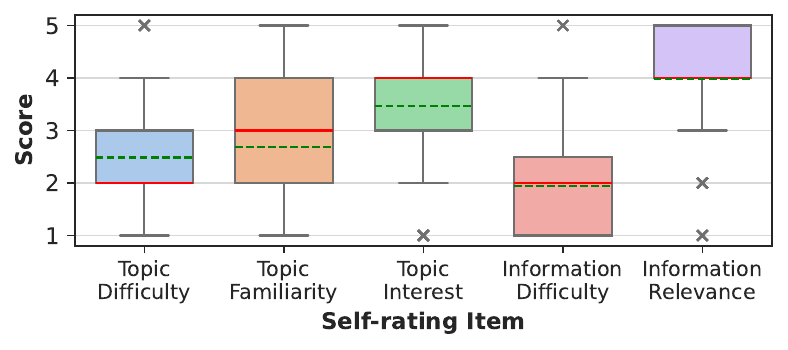}
    \caption{Distribution of Self-rated Task Perceptions on Perceived Difficulty, Familiarity, Interest in the Topic, and Perceived Difficulty and Relevance on the received Information.}
    \Description{A bar chart displaying participants’ ratings of topic difficulty, familiarity, interest, and perceived relevance and difficulty of the search results. It confirms that the stimuli met the expected criteria.}
    \label{fig:self-ratings}    
\end{figure}

As summarized in Figure~\ref{fig:self-ratings}, the participants have rated Topic Difficulty as 2.5 (SD: 1.1), Topic Familiarity as 2.7 (SD: 1.2),
Topic Interest is 3.5 (SD: 1.1). For the search results, the participants have rated Information Difficulty as 1.9 (SD: 1.0) and Information Relevance as 4.0 (SD: 1.1). Overall, the task perceptions have met our anticipation, all data have been processed further for analysis.

\paragraph*{Task Durations}

As reported in Table~\ref{tab:dataset_size_full}, the participants 
take an average of {23.4 (SD: 14.1)} seconds to complete reading the backstories at \ifn, and 10 seconds (as pre-defined) to think about a query at \qf.
The interaction modality affects completion time. \qss takes an average of 6.8 (SD: 2.5) seconds, faster than \qst, which takes an average of 17.7 (SD: 10.1) seconds. \rjl takes an average of 66.0 (SD: 4.4) seconds, longer than \rjr, which takes an average of 49.2 (SD: 19.8) seconds.

\subsection{Understanding Physiological Responses at Different Stages of Information Search}
\label{subsec:responses_stages}

\paragraph*{Brain Activity Responses}

\begin{table}[tp]
\caption{{Non-parametric Friedman Statistical Testing Results (F-value) comparing all search stages and \eyeopen for the 4 frequency bands at 14 EEG nodes. $p<.05$*, $<.01$**, $< .001$***. }}
\label{tab:statistic_testings_eeg}
\sisetup{
    table-format = 2.3,
    detect-weight = true,
    table-space-text-post = ***
}
\resizebox{0.99\textwidth}{!}{
\begin{tabular}{l S[table-format=2.3@{\,}] S[table-format=2.3@{\,}] S[table-format=2.3@{\,}] S[table-format=2.3@{\,}] S[table-format=2.3@{\,}]
                S[table-format=2.3@{\,}] S[table-format=2.3@{\,}] S[table-format=2.3@{\,}] S[table-format=2.3@{\,}] S[table-format=2.3@{\,}]
                S[table-format=2.3@{\,}] S[table-format=2.3@{\,}] S[table-format=2.3@{\,}] S[table-format=2.3@{\,}]}
    \toprule
    \rowcolor[HTML]{EFEFEF} 
    \textbf{Band\textbackslash Node} & \textbf{AF3} & \textbf{AF4} & \textbf{FC5} & \textbf{FC6} & \textbf{F3} & \textbf{F4} & \textbf{F7} & \textbf{F8} & \textbf{T7} & \textbf{T8} & \textbf{P7} & \textbf{P8} & \textbf{O1} & \textbf{O2} \\ 
    \midrule
    \textbf{Theta}      & 1.850      & 3.052{**}    & 0.737       & 5.387{***}   & 2.366{*}    & 7.667{***}  & 2.569{*}    & 2.648{*}    & 4.070{**}   & 3.034{**}   & 8.366{***}  & 6.261{***}  & 7.441{***}  & 11.601{***} \\
    \textbf{Alpha}      & 8.687{***} & 11.116{***}  & 15.794{***} & 13.900{***}  & 13.870{***} & 29.763{***} & 11.948{***} & 17.068{***} & 23.055{***} & 29.342{***} & \underline{32.476}{***} & 23.458{***} & 28.649{***} & 15.345{***} \\
    \textbf{Beta}       & 10.223{***}& 10.621{***}  & 12.942{***} & 8.465{***}   & 7.047{***}  & 9.358{***}  & 15.278{***} & 14.554{***} & 8.508{***}  & 7.991{***}  & 12.368{***} & 12.479{***} & 11.850{***} & 9.208{***}  \\
    \textbf{LowGamma}   & 5.848{***} & 8.949{***}   & 9.945{***}  & 6.160{***}   & 5.062{***}  & 4.535{***}  & 9.850{***}  & 8.254{***}  & 2.745{*}    & 4.750{***}  & 1.411       & 3.720{**}   & 5.116{***}  & 10.531{***} \\
    \bottomrule
\end{tabular}

}
\end{table}

\begin{figure}[htbp!]
    \centering
    \includegraphics[width=0.7\linewidth]{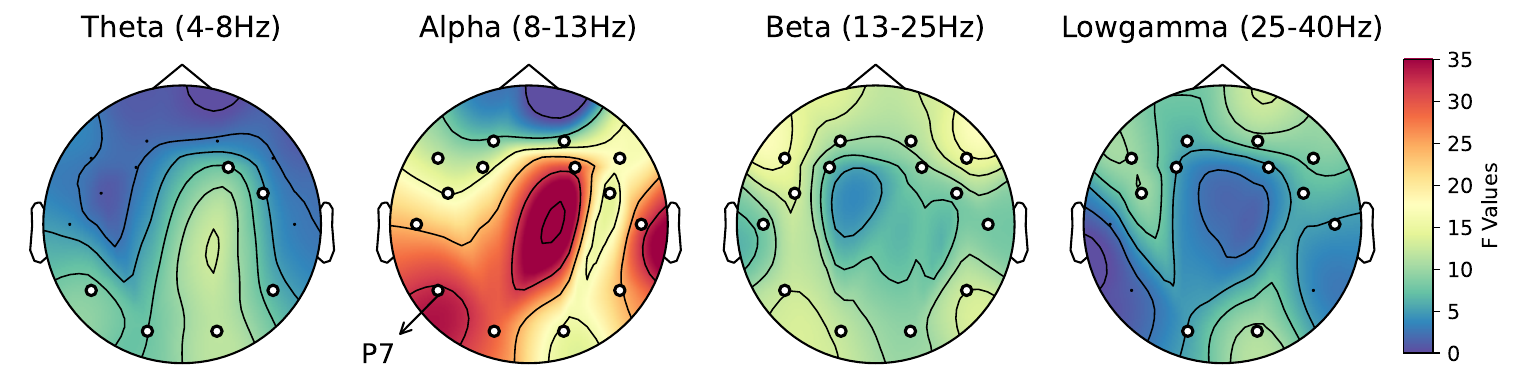}
    \caption{Topography which shows the significance of difference (F-value from Friedman Test) between brain response in the baseline, \eyeopen and the 6 search stages (\ifn, \qf, \qst, \qss, \rjr, \rjl). Highlighted channels indicate the differences are significant at $p<.001$. Greenhouse-Geisser correction is applied to adjust for the violations of sphericity. }
    \Description{Topography heatmap showing EEG channel-wise statistical differences across search stages for four frequency bands (Theta, Alpha, Beta, Low Gamma). Highlighted regions indicate significant cognitive activity variations.} 
    \label{fig:eeg-6-topo}
\end{figure}

\begin{table}[tp]
\caption{Pairwise Wilcoxon Test Results for the EEG P7 Alpha bands. $p< .01$**, $< .001$***. One-step Bonferroni Correction applied for multiple comparisons. }
\label{tab:statistic_testings_p7}
\resizebox{0.7\textwidth}{!}{
\begin{tabular}{@{}cccccccccccc@{}}
\toprule
\rowcolor[HTML]{EFEFEF} 
\textbf{A}          & \multicolumn{6}{c}{\textbf{\eyeopen}}         & \multicolumn{5}{c}{\textbf{\ifn}}  \\ \cmidrule(lr){2-7}  \cmidrule(lr){8-12} 
\textbf{B}          & \textbf{\ifn}          & \textbf{\qf}           & \textbf{\qss}         & \textbf{\qst} & \textbf{\rjr} & \textbf{\rjl}          & \textbf{\qf}           & \textbf{\qss}         & \textbf{\qst}          & \textbf{\rjr}   & \textbf{\rjl}           \\ \midrule
\textbf{W}          & 45                     & 3                      & 39                    & 65            & 32            & 4                      & 7                      & 14                    & 8                      & 65              & 9                       \\
\textit{\textbf{p}} & {.503}             & \textbf{.000***} & {.254}            & {1}       & {.102}    & \textbf{{.000}***} & \textbf{{.001}**}  & \textbf{{.004}**} & \textbf{{.001}**}  & {1}         & \textbf{{.001}**}   \\
\bottomrule 
\toprule
\rowcolor[HTML]{EFEFEF} 
\textbf{A}          & \multicolumn{4}{c}{\textbf{\qf}}                                                        & \multicolumn{3}{c}{\textbf{\qss}}                               & \multicolumn{2}{c}{\textbf{\qst}}              & \multicolumn{2}{c}{\textbf{\rjr}}         \\  \cmidrule(lr){2-5} \cmidrule(lr){6-8} \cmidrule(lr){9-10} \cmidrule(lr){11-12}
\textbf{B}          & \textbf{\qss}          & \textbf{\qst}          & \textbf{\rjr}         & \textbf{\rjl} & \textbf{\qst} & \textbf{\rjr}          & \textbf{\rjl}          & \textbf{\rjr}         & \textbf{\rjl}          & \multicolumn{2}{c}{\textbf{\rjl}}         \\ \midrule
\textbf{W}          & 0                      & 1                      & 16                    & 103           & 65            & 10                     & 0                      & 10                    & 1                      & \multicolumn{2}{c}{16}                    \\
\textit{\textbf{p}} & \textbf{{.000}***} & \textbf{{.000}***} & \textbf{{.007}**} & {1}       & {1}       & \textbf{{.002}**}  & \textbf{{.000}***} & \textbf{{.002}**} & \textbf{{.000}***} & \multicolumn{2}{c}{\textbf{{.007}**}} \\ \bottomrule
\end{tabular}
}
\end{table}

As presented in Figure~\ref{fig:eeg-6-topo},
the $Alpha$ bands show the most prominent differences, with the highest F-values concentrated in the left-parietal, i.e., $P7$ ($F[6, {112}] = {32.476}, p<.001$), left-occipital regions, i.e., $O1$ ($F[6, {112}] = {28.649}, p<.001$), and right-frontal regions, i.e., $F4$ ($F[6, 112] = 29.763, p<.001$). 
The differences observed in $Alpha$ band likely reflect task-related variations in attention allocation and spatial processing~\cite{randall2014putting, moravec2018fake}, e.g., for filtering out the irrelevant information\cite{ye2023relevance}, across different information stages.
The $Beta$ bands also show significant differences across all regions. As $Beta$ activity is commonly associated with active thinking and problem solving~\cite{KUMAR20122525, yu2023people}, the results suggest that these processes might be similarly and significantly engaged across all conditions. 
Lastly, the $Theta$ and $LowGamma$ bands have generally lower F-values.  
This might indicate that both $Theta$ and $LowGamma$ are less sensitive to the specific differences among the search stages compared to the other bands.

\paragraph*{Pupil Dilation}

\begin{figure}[htbp!]
    \centering
    \includegraphics[width=0.4\linewidth]{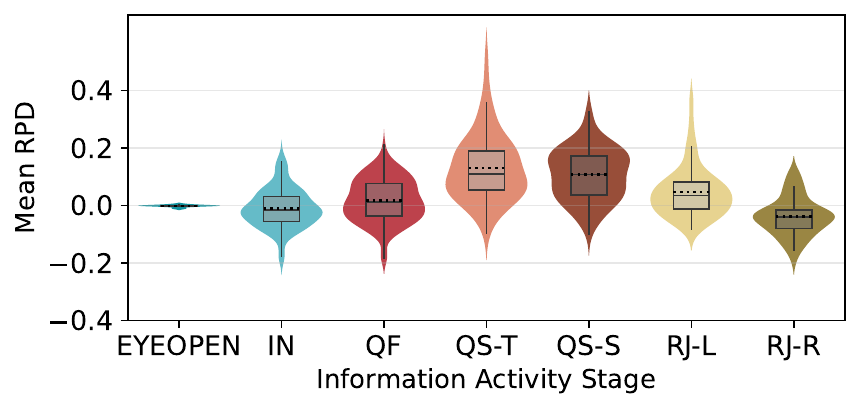}
    \caption{Violin plot represents the distribution of mean Relative Pupil Dilation (RPD) across the baseline \eyeopen and the 6 search stages (\ifn, \qf, \qst, \qss, \rjr, \rjl).}
    \Description{A violin plot comparing mean pupil dilation across different search stages and baseline. It visualizes cognitive load and attention levels during each stage.}
    \label{fig:rpd-violin}
\end{figure}

In a constantly illuminated condition, Relative Pupil Dilation (RPD) can indicate attention and exerted cognitive load \cite{Gwizdka2017temporal}. 
As demonstrated in Figure~\ref{fig:rpd-violin}, both modalities for \qs show larger RPD compared to the other stages. The Friedman test revealed significant difference ($F[6, {100}] = {37.962}, p<.001$) among search stages (refer to Table~\ref{tab:statistic_testings}).
And the post-hoc test reveals that both \qss and \qst are significantly different from \eyeopen and all other stages. In particular, $W(\textrm{\eyeopen}, \textrm{\qss}) = 4, p<.01$, $W(\textrm{\eyeopen}, \textrm{\qst}) = 1, p<.001$, $W(\textrm{\ifn}, \textrm{\qss}) = 0, p<.001$, $W(\textrm{\ifn}, \textrm{\qst}) = 0, p<.001$,
$W(\textrm{\qf}, \textrm{\qss}) = 0, p<.001$, $W(\textrm{\qf}, \textrm{\qst}) = 3, p<.01$,
$W(\textrm{\rjr}, \textrm{\qss}) = 0, p<.001$, $W(\textrm{\rjr}, \textrm{\qst}) = 0, p<.001$,
$W(\textrm{\rjl}, \textrm{\qst}) = 19, p<.05$.
This result might indicate the demanding efforts for formulating and expressing queries in text or speech to the external system \cite{shovon2015search}.

\paragraph*{Peripheral Responses}
Statistical testing results in Table~\ref{tab:statistic_testings} found no significant difference in the mixed or tonic EDA values (SCL) . Yet, a significant difference is revealed in phasic values, SCR ($F[6, {112}] = {5.259}, p<.001$).

\begin{table}[tp]
\caption{{Friedman Statistical Testing Results comparing among all search stages and \eyeopen. $p<.001$***.}}
\label{tab:statistic_testings}
\centering
\sisetup{
    detect-weight = true,
    table-space-text-post = ***,
        add-integer-zero=false
}

\resizebox{0.46\textwidth}{!}{
\begin{tabular}{@{}llSS}
\toprule
\rowcolor[HTML]{EFEFEF} 
\textbf{Data Type} & \textbf{Feature} & \textbf{F}& \textit{\textbf{p}}     \\ \midrule
\textbf{EDA} [6,112]& Mean Mixed EDA        & 1.321 & .254 \\
                   & Mean SCL        & 1.244 & .290 \\ 
                   & Mean SCR
                   & .259 & \bfseries.000{***} \\ 
                   \midrule
\textbf{PUPIL} [6, 100]& Mean RPD        & 37.962 & \bfseries .000{***} \\
\bottomrule
\end{tabular}}
\end{table}

\subsection{The influence of Cognitive Intents:
Comparing Eye-Driven Activities}
\label{subsec:influence_intents}

\begin{figure}[htbp!]
    \centering
    \begin{subfigure}[t]{0.53\linewidth}
        \centering
        \includegraphics[width=\linewidth]{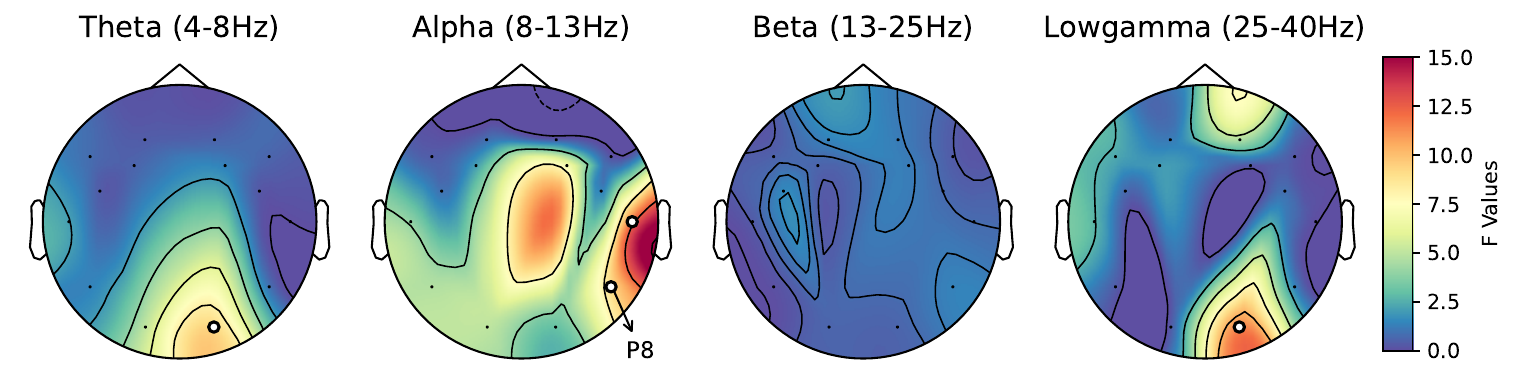}
        \caption{{Friedman test results of EEG $Alpha$ band. Highlighted channels indicate the differences are significant at $p < .001$.}}
        \label{subfig:eeg-in-ec-read-topo}
    \end{subfigure}
    \hspace{10pt}
    ~
    \begin{subfigure}[t]{0.43\linewidth}
        \centering
        \includegraphics[width=0.7\linewidth]{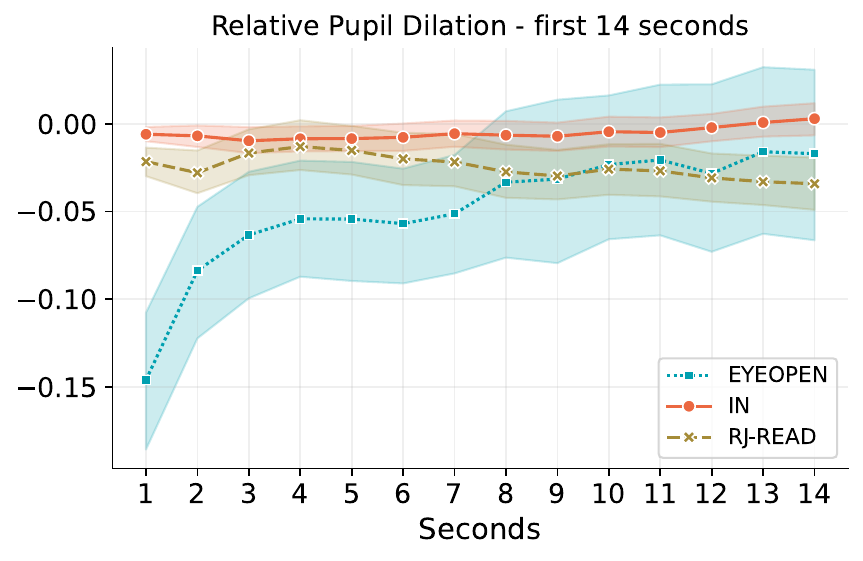}
        \caption{{Changes in Relative Pupil Diameter (RPD) during the first 15 seconds following stage onset. The values are subtracted from the value at the first second.}}
        \label{subfig:eeg-in-ec-read-pupil}
    \end{subfigure}
    \vspace{-8pt}
    \caption{The physiological responses over 3 eye-driven activities with different intentions, \eyeopen, \ifn, and \rjr.}
    \Description{(a) EEG topography showing differences in Alpha band activity across EYEOPEN, IN, and RJ-R stages. (b) Line graph showing changes in pupil dilation over time during these stages.}
    \label{fig:rpd-4}
\end{figure}

The stages, \eyeopen, \ifn, and \rjr, require visual engagement but vary in cognitive demands and intentionality. 
As demonstrated in Figure~\ref{subfig:eeg-in-ec-read-topo}, the Friedman test reveals significant differences in the temporal-parietal-occipital regions on the right hemisphere among the 3 stages.
This suggests that cognitive demands or the focus of attention differ primarily in these areas. 
In particular, the post-hoc Wilcoxon test reveals that
the $Alpha$ band during \eyeopen is significantly different than during \ifn at T8, $W(\textrm{\eyeopen}, \textrm{\ifn}) = {9}, p<{.01}$. When comparing to \rjr, the difference found at T8, $W(\textrm{\eyeopen}, \textrm{\rjr}) = {11}, p<{.01}$, and P8, $W(\textrm{\eyeopen}, \textrm{\rjr}) = {26}, p<{.05}$. Besides, the $Theta$ bands also found significant differences at occipital regions. More specifically, $Theta$ at O1 differs between \eyeopen and \rjr, $W(\textrm{\eyeopen}, \textrm{\rjr}) = 27, p<.05$, and at O2 differs between \ifn and \rjr, $W(\textrm{\ifn}, \textrm{\rjr}) = 9, p<.01$.

{The mean RPD at \eyeopen does not significantly differ from \ifn, $W(\textrm{EYEOPEN}, \textrm{\ifn}) = {63}, p>.05$, or \rjr $W(\textrm{EYEOPEN}, \textrm{\rjr}) = {35}, p>.05$. But it is significantly different between \ifn and \rjr, $W(\textrm{\ifn}, \textrm{\rjr}) = {1}, p <.001$.}  
Figure~\ref{subfig:eeg-in-ec-read-pupil} illustrates the changes in pupillary responses during the first 15 seconds (align with the shortest stage, \eyeopen), normalized to the average value of the onset second.
At \eyeopen, as participants stared at the screen, their pupils dilated overall, reflecting heightened attention and cognitive engagement. During \ifn, the pupillary responses remained relatively stable with a slight increase. This suggests the participants might sustain attention or consistent cognitive processing to comprehend and recognize their task. In contrast, at \rjr, the pupil diameters generally contracted with a convex response at the beginning. This might imply a brief period of heightened attention or arousal, possibly due to curiosity or interest when the search result first appeared, followed by a gradual decline.

\subsection{The Influence of Interaction Modalities}
\label{subsec:influence_modalities}

\begin{figure}[htbp!]
    \centering
    \begin{subfigure}[t]{0.47\linewidth}
        \includegraphics[width=\linewidth]{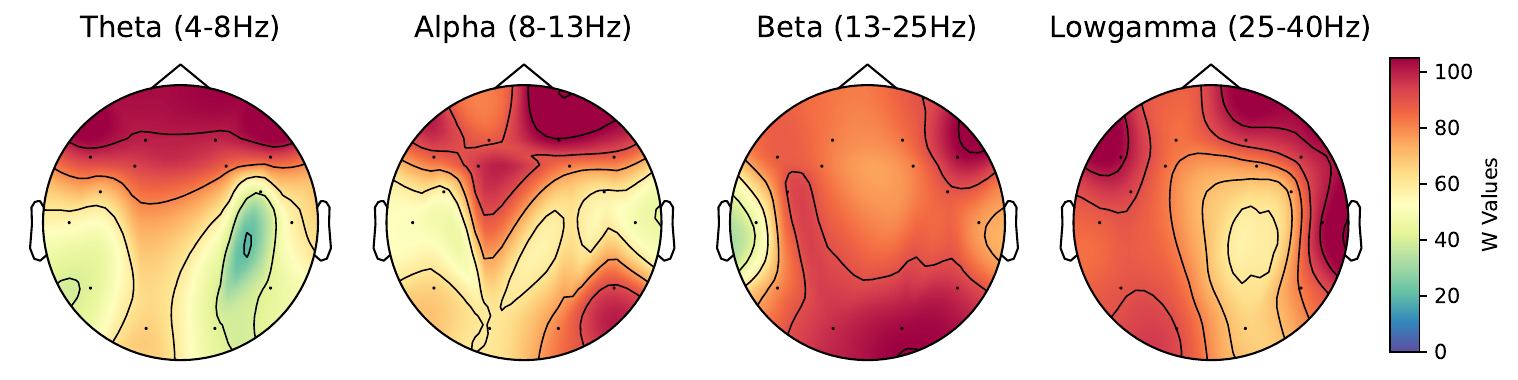}
        \caption{{\qss v.s. \qst}}
        \label{subfig:eeg-speak-type-topo}
    \end{subfigure}
    ~
    \begin{subfigure}[t]{0.47\linewidth}
        \includegraphics[width=\linewidth]{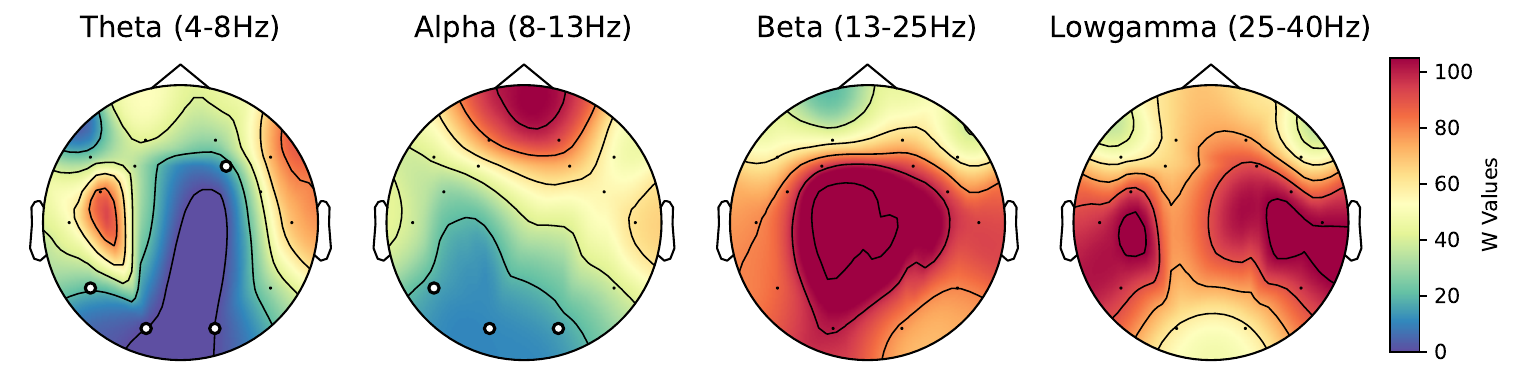}
        \caption{{\rjr v.s. \rjl}}
        \label{subfig:eeg-read-listen-topo}
    \end{subfigure}
     \vspace{-8pt}
    \caption{Topography which shows the significance of difference (W-value) between brain responses between interaction modalities. Bonferroni Correction applied. Highlighted channels indicate the differences are significant at $p<.05$.}
    \Description{Topographic maps comparing EEG responses between typing vs. speaking (QS-T vs. QS-S) and reading vs. listening (RJ-R vs. RJ-L). Significant differences are marked.}
    \label{fig:eeg-interaction-modal}
\end{figure}

As presented in Figure~\ref{subfig:eeg-speak-type-topo}, no significant difference is found between \qst and \qss for query submission. And the RPD shows no significant difference, $W(\textrm{\qss}, \textrm{\qst}) = 21, p>.05$. 

Comparing between \rjr and \rjl, the differences are observed mainly at the parietal-occipital regions {(i.e., P7, O1, O2)} for both $Theta$ and $Alpha$ bands (refer to Figure~\ref{subfig:eeg-read-listen-topo}): $W(\textrm{P7}, \textrm{Theta}) = {22}, p<.0{5}$, $W(\textrm{P7}, \textrm{Alpha}) = {16}, p<.01$, $W(\textrm{O1}, \textrm{Theta}) = {0}, p<.001$, $W(\textrm{O1}, \textrm{Alpha}) = {11}, p<.01$,  $W(\textrm{O2}, \textrm{Theta}) = {0}, p<.001$, $W(\textrm{O2}, \textrm{Alpha}) = {16}, p<.01$. 
There is also a significant difference observed at the frontal region for $Theta$ band, $W(\textrm{F4}) = {15}, p<.01$.
Besides, the RPD shows a significant difference, $W(\textrm{\rjr}, \textrm{\rjl}) = 1, p<.001$.

\subsection{Classification Performance of the Sensor Data}
\label{sec:classification}

\subsubsection{Experimental Setting}

The objective is to classify all search stages in considering interaction modalities, \ifn, \qf, \qss, \qst, \rjr, and \rjl. 
We categorized the sensor data into 2 types -- physiological and behavioral data -- and 5 groups -- including EDA, EEG, PUPIL, GAZE, and MOTION. MOTION combines the wrist and head motion.

\paragraph{Classifiers}
The classification experiment was developed with the Scikit-learn Python Library. 
The support vector machine (SVM) with an RBF kernel was used as the base model \cite{so2017evaluation, bias2023boon}.
We have built two types of models, one with only features from every single type of sensor data (results refer to Section~\ref{subsec:single-sensor}), the other was a late-decision-fusion model which combines two or more trained single-sensor models (results refer to Section~\ref{subsec:descision-fusion}). We ensemble the model with a soft voting classifier with equal weights assigned to the base models~\cite{koelstra2011deap}. The soft voting classifier calculates the $\argmax$ of the sums of the predicted probabilities computed by the base models.

\paragraph{Dimension Reduction \& Hyper-Parameter Tuning}

Studies involving multiple physiological data often result in a large number of extracted features \cite{koelstra2011deap, Gwizdka2017temporal}, necessitating dimension reduction for the subsequent process, e.g., model training. We implemented with GridSearchCV from the Scikit-learn library, and tuned the model with different dimension reduction techniques, including Linear Discriminant Analysis, Neighborhood Components Analysis, and Select-K-Best (using mutual information or F-statistic), and parameters including the regularization parameter (C), kernel coefficient (gamma), and polynomial degree.

\paragraph{Evaluation}
We conducted Leave-One-participant-Out (LOO) cross-validation and reported accuracy and F1-macro metrics for the evaluation. LOO cross-validation assesses the performance of a model by iteratively training on data from all participants except one, which is reserved for testing. In this project, which involves data from 20 participants, each model was evaluated 20 times, with the data from 19 participants used for training and the data from 1 participant used for testing each time\footnote{Note that as 2 participants are missing the eye-tracking data, the models involving PUPIL or GAZE were evaluated 18 times.}. 
Accuracy reflects the overall correctness of a model's predictions, while the F1-macro score addresses class imbalance by averaging performance equally across all classes, offering a balanced evaluation of model effectiveness. Higher scores for both metrics indicate better model performance. We employed the random guess baseline, which is around 17\%.

\subsubsection{Single-Sensor Performance}
\label{subsec:single-sensor}

As presented in Table~\ref{tab:classification-results-6class} and Figure~\ref{fig:auc-single}, PUPIL consistently demonstrated the best performance, while EDA was the only model to perform below baseline on F1 score.
When comparing window feature settings, we observed that 1--0 and 2--1 windows produced identical performance metrics across all sensor groups, suggesting minimal impact of this particular window size variation. However, implementing the 4--2 window features yielded variable effects across sensor groups. GAZE showed the most substantial improvement with a 10.8\% increase in accuracy and 10.0\% increase in F1. PUPIL and MOTION demonstrated moderate enhancements (4.4\% increase in both metrics for PUPIL; 5.4\% accuracy and 2.5\% F1 increase for MOTION), while EEG showed minimal improvement (1.1\% increase in accuracy, 0.6\% increase in F1). Interestingly, EDA exhibited a decrease of 3.6\% in F1 score.

Figure~\ref{fig:cm-prediction-42} reveals distinct classification patterns across models. The EDA model showed poor discriminative capability with predictions skewed toward \ifn and \qst. While EEG provided more balanced predictions, it still favored \qst and \rjl. Only the PUPIL model {produced} well-balanced and accurate predictions across all classes, while GAZE and MOTION models exhibited biases similar to EDA.
Notably, when comparing the eye-driven activities \ifn and \rjr (which involve different cognitive intentions), only MOTION and PUPIL models successfully differentiated between them. This suggests that similar gaze movement patterns in reading activities may confound the GAZE model, while the wristband may lack sufficient sensitivity to capture subtle EDA variations during these activities.

\begin{figure}[htbp!]
    \centering
    \includegraphics[width=.85\linewidth]{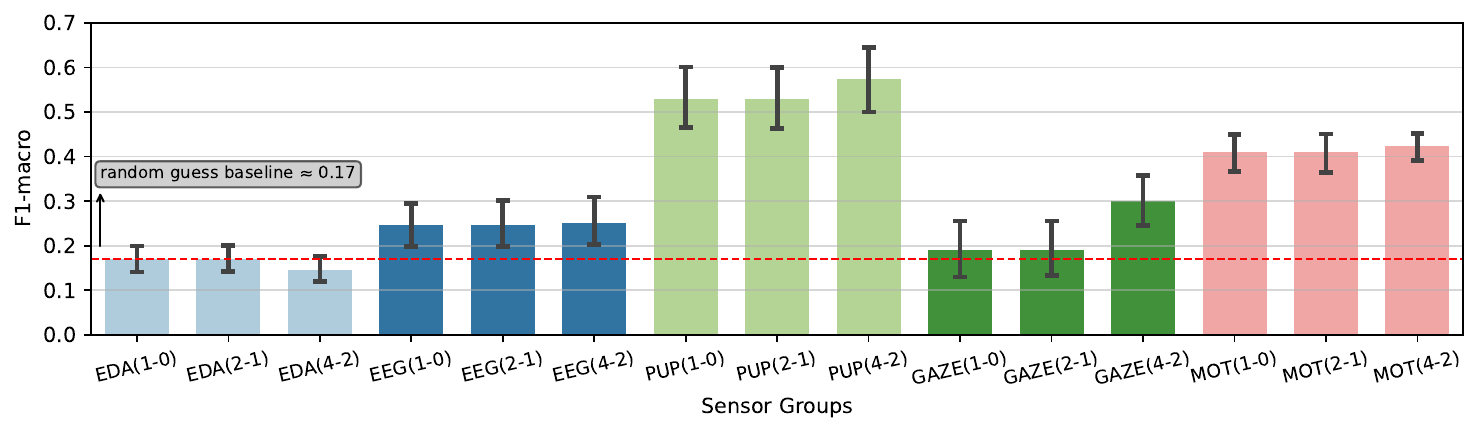}
     \caption{The performance in F1-macro metrics of the SVM models for single sensor groups and different time-window features. The sensor groups are EDA, EEG, PUPIL, GAZE, and MOTION (from left to right).  Error bars present 95\% confidence intervals.}
     \Description{A bar chart comparing classification performance (F1-macro) of models using different sensor types and time windows. PUPIL and GAZE show the highest scores.}
    \label{fig:auc-single}
\end{figure}

\begin{table}[tp]
\caption{Classification Performance on SVM with RBF kernel of single sensor groups using features extracted from different time windows. The evaluation metrics are Accuracy and F1-macro (with standard deviation).}
\label{tab:classification-results-6class}
\resizebox{0.95\textwidth}{!}{
\begin{tabular}{@{}c|cc|cc|cc|cc|cc@{}}
\toprule
\rowcolor[HTML]{EFEFEF} 
\textbf{Sensor Group} & \multicolumn{2}{c|}{\textbf{EDA}} & \multicolumn{2}{c|}{\textbf{EEG}} & \multicolumn{2}{c|}{\textbf{PUPIL}} & \multicolumn{2}{c|}{\textbf{MOTION}} & \multicolumn{2}{c}{\textbf{GAZE}} \\ 
\rowcolor[HTML]{EFEFEF} 
Feature Window  & Accuracy   & F1-macro   & Accuracy   & F1-macro   & Accuracy     & F1-macro    & Accuracy   & F1-macro  & Accuracy   & F1-macro   \\  \midrule
 1s--0s overlap& 0.273 (0.065)& 0.170 (0.067)& 0.324 (0.092)& 0.246 (0.116)& 0.549 (0.145)& 0.530 (0.153)& 0.452 (0.114)& 0.409 (0.098)& 0.295 (0.124)&0.191 (0.139)\\
2s--1s overlap & 0.273 (0.065)& 0.170 (0.067)& 0.324 (0.092)& 0.246 (0.116)& 0.549 (0.145)& 0.530 (0.153)& 0.452 (0.114)& 0.409 (0.098)& 0.295 (0.124)&  0.191 (0.139)\\
4s--2s overlap & 0.282 (0.067)& 0.146 (0.067)& 0.328 (0.118)& 0.252 (0.128)& 0.585 (0.151)& 0.574 (0.152)& 0.506 (0.077)& 0.424 (0.071)& 0.403 (0.104)&  0.299 (0.126)\\ \bottomrule
\end{tabular}
}
\end{table}

\begin{figure}[htbp!]
    \centering
    \includegraphics[width=0.9\linewidth]{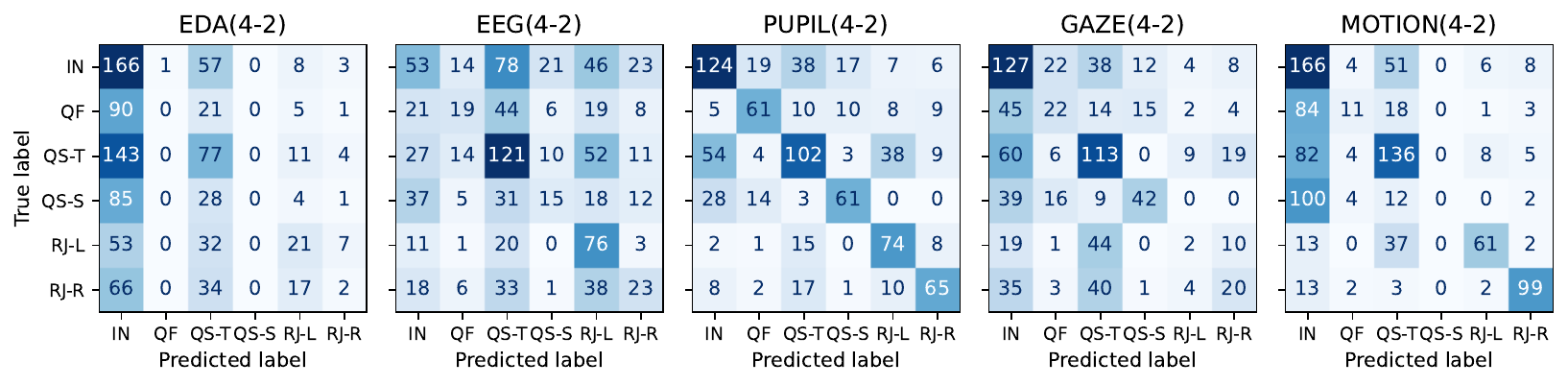}
    \caption{Confusion Matrix of single sensor groups with 4 seconds with 2-second overlap time window features.}
    \Description{Confusion matrices for each sensor type showing how well each model classified the six search stages. PUPIL and MOTION models show more balanced predictions.}
    \label{fig:cm-prediction-42}
\end{figure}

\subsubsection{Decision-Fusion Performance}
\label{subsec:descision-fusion}

\setlength{\arrayrulewidth}{0.5pt}
\begin{table}[tb]
\caption{Classification Performance of SVM with RBF kernel of fusion models and 4-2 window features. The highest and lowest in terms of F1 scores are highlighted in bold text. The evaluation metrics are Accuracy and F1-macro (with standard deviation).}
\label{tab:fusion-results}
\resizebox{0.99\textwidth}{!}{
\begin{tabular}{|>{\columncolor{lightgray}}r|c|c|c|c|l|c|ccccclc}

\hhline{|-|-|-|-|-|-|-|~|-|-|-|-|-|-|}
 & \cellcolor{lightgray}\textbf{EEG}                     & \cellcolor{lightgray}\textbf{PUPIL}                   & \cellcolor{lightgray}\textbf{MOTION}                  & \cellcolor{lightgray}\textbf{GAZE}                    & \multicolumn{1}{c|}{\cellcolor{lightgray}\textbf{Accuracy}} & \cellcolor{lightgray}\textbf{F1}         &     & \multicolumn{1}{|c|}{\cellcolor{lightgray}}
 & \multicolumn{1}{c|}{\cellcolor{lightgray}\textbf{PUPIL}} & \multicolumn{1}{c|}{\cellcolor{lightgray}\textbf{MOTION}} & \multicolumn{1}{c|}{\cellcolor{lightgray}\textbf{GAZE}} & \multicolumn{1}{c|}{\cellcolor{lightgray}\textbf{Accuracy}} & \multicolumn{1}{c|}{\cellcolor{lightgray}\textbf{F1}} \\ \cline{2-7} \cline{10-14} 
 
   & \color[HTML]{009901}\checkmark & \color[HTML]{009901}\checkmark & \color[HTML]{009901}\checkmark & \color[HTML]{009901}\checkmark & 0.637 (0.143)                    & 0.609
 (0.158) & & \multicolumn{1}{|c|}{\cellcolor{lightgray}}     & \multicolumn{1}{c|}{\color[HTML]{009901}\checkmark}       & \multicolumn{1}{c|}{\color[HTML]{009901}\checkmark}        & \multicolumn{1}{c|}{\color[HTML]{009901}\checkmark}      & \multicolumn{1}{l|}{\textbf{0.671 (0.147)}}                    & \multicolumn{1}{c|}{\textbf{0.645 (0.161)}}              \\ \cline{2-7} \cline{10-14} 

 & \color[HTML]{009901}\checkmark & \color[HTML]{009901}\checkmark & \color[HTML]{009901}\checkmark & \color[HTML]{FE0000} --          & 0.570 (0.115)                                                  & 0.525 (0.129)                         &      & \multicolumn{1}{|c|}{\cellcolor{lightgray}}                               & \multicolumn{1}{c|}{\color[HTML]{009901}\checkmark}       & \multicolumn{1}{c|}{\color[HTML]{009901}\checkmark}        & \multicolumn{1}{c|}{\color[HTML]{FE0000} --}               & \multicolumn{1}{l|}{0.551 (0.146)}                             & \multicolumn{1}{c|}{0.508 (0.155)}                       \\ \cline{2-7} \cline{10-14} 

  & \color[HTML]{009901}\checkmark & \color[HTML]{009901}\checkmark & \color[HTML]{FE0000} --          & \color[HTML]{009901}\checkmark & 0.552 (0.159)                                                  & 0.519 (0.159)                       &        & \multicolumn{1}{|c|}{\cellcolor{lightgray}}                               & \multicolumn{1}{c|}{\color[HTML]{009901}\checkmark}       & \multicolumn{1}{c|}{\color[HTML]{FE0000} --}                 & \multicolumn{1}{c|}{\color[HTML]{009901}\checkmark}      & \multicolumn{1}{l|}{0.570 (0.154)}                             & \multicolumn{1}{c|}{0.534 (0.164)}                       \\ \cline{2-7} \cline{10-14} 

   & \color[HTML]{009901}\checkmark & \color[HTML]{FE0000} --          & \color[HTML]{009901}\checkmark & \color[HTML]{009901}\checkmark & 0.569 (0.170)                                                  & 0.537 (0.184)                           &    & \multicolumn{1}{|c|}{\cellcolor{lightgray}}                               & \multicolumn{1}{c|}{\color[HTML]{FE0000} --}                & \multicolumn{1}{c|}{\color[HTML]{009901}\checkmark}        & \multicolumn{1}{c|}{\color[HTML]{009901}\checkmark}      & \multicolumn{1}{l|}{0.556 (0.165)}                             & \multicolumn{1}{c|}{0.531 (0.164)}                       \\ \cline{2-7} \cline{10-14} 

  & \color[HTML]{FE0000} --          & \color[HTML]{009901}\checkmark & \color[HTML]{009901}\checkmark & \color[HTML]{009901}\checkmark & 0.647 (0.147)                                                  & 0.614 (0.145)                       &        & \multicolumn{1}{|c|}{\cellcolor{lightgray}}                               & \multicolumn{1}{c|}{\color[HTML]{009901}\checkmark}       & \multicolumn{1}{c|}{\color[HTML]{FE0000} --}                 & \multicolumn{1}{c|}{\color[HTML]{FE0000} --}               & \multicolumn{1}{l|}{0.555 (0.133)}                             & \multicolumn{1}{c|}{0.533 (0.137)}                       \\ \cline{2-7} \cline{10-14} 

       & \color[HTML]{009901}\checkmark & \color[HTML]{009901}\checkmark & \color[HTML]{FE0000} --          & \color[HTML]{FE0000} --          & 0.535 (0.141)                                                  & 0.499 (0.147)                       &        & \multicolumn{1}{|c|}{\cellcolor{lightgray}}                               & \multicolumn{1}{c|}{\color[HTML]{FE0000} --}                & \multicolumn{1}{c|}{\color[HTML]{009901}\checkmark}        & \multicolumn{1}{c|}{\color[HTML]{FE0000} --}               & \multicolumn{1}{l|}{0.452 (0.114)}                             & \multicolumn{1}{c|}{0.396 (0.121)}                       \\ \cline{2-7} \cline{10-14} 

 & \color[HTML]{009901}\checkmark & \color[HTML]{FE0000} --          & \color[HTML]{009901}\checkmark & \color[HTML]{FE0000} --          & 0.475 (0.108)       & 0.405 (0.107)     &       & 
\multicolumn{1}{|c|}{\multirow{-7}{*}{\cellcolor{lightgray}\textbf{EEG}}} & \multicolumn{1}{c|}{\color[HTML]{FE0000} --}                & \multicolumn{1}{c|}{\color[HTML]{FE0000} --}                 & \multicolumn{1}{c|}{\color[HTML]{009901}\checkmark}      & \multicolumn{1}{l|}{0.490 (0.124)}                             & \multicolumn{1}{c|}{0.430 (0.123)}                       
\\ \cline{2-7} \cline{9-14} 

 & \color[HTML]{009901}\checkmark & \color[HTML]{FE0000} --          & \color[HTML]{FE0000} --          & \color[HTML]{009901}\checkmark & 0.525 (0.135)                                                  & 0.458 (0.148)  & &    & &  &   &   & \\  \hhline{|~|-|-|-|-|-|-|~|-|-|-|-|-|-|}

       & \color[HTML]{FE0000} --          & \color[HTML]{009901}\checkmark & \color[HTML]{009901}\checkmark & \color[HTML]{FE0000} --          & 0.598 (0.154)                   & 0.580 (0.139)         
& & \multicolumn{2}{|c|}{\cellcolor{lightgray}}    
& \multicolumn{1}{c|}{\cellcolor{lightgray}\textbf{MOTION}} & \multicolumn{1}{c|}{\cellcolor{lightgray}\textbf{GAZE}} & \multicolumn{1}{c|}{\cellcolor{lightgray}\textbf{Accuracy}} & \multicolumn{1}{c|}{\cellcolor{lightgray}\textbf{F1}} 
\\ \cline{2-7} \cline{11-14}

\cellcolor{lightgray}                                & \color[HTML]{FE0000} --          & \color[HTML]{009901}\checkmark & \color[HTML]{FE0000} --          & \color[HTML]{009901}\checkmark & 0.591 (0.146)                                                  & 0.550 (0.150)     & &
\multicolumn{2}{|c|}{\cellcolor{lightgray}}    & \multicolumn{1}{c|}{\color[HTML]{009901}\checkmark}        & \multicolumn{1}{c|}{\color[HTML]{009901}\checkmark}      & \multicolumn{1}{c|}{0.645 (0.153)}                             & \multicolumn{1}{c|}{0.621 (0.146)}   
\\ \cline{2-7} \cline{11-14} 

& \color[HTML]{FE0000} --          & \color[HTML]{FE0000} --          & \color[HTML]{009901}\checkmark & \color[HTML]{009901}\checkmark & 0.543 (0.114)                             & 0.486 (0.124)  &  &
\multicolumn{2}{|c|}{\cellcolor{lightgray}}                  & \multicolumn{1}{c|}{\color[HTML]{009901}\checkmark}        & \multicolumn{1}{c|}{\color[HTML]{FE0000} --}               & \multicolumn{1}{c|}{0.597 (0.111)}                             & \multicolumn{1}{c|}{0.590 (0.119)}   
\\ \cline{2-7} \cline{11-14} 

  & \color[HTML]{009901}\checkmark & \color[HTML]{FE0000} --          & \color[HTML]{FE0000} --          & \color[HTML]{FE0000} --          & \textbf{0.329 (0.090)}                                         & \textbf{0.234 (0.099)} &   &
\multicolumn{2}{|c|}{\multirow{-3}{*}{\cellcolor{lightgray}\textbf{PUPIL}}}                                                             & \multicolumn{1}{c|}{\color[HTML]{FE0000} --}                 & \multicolumn{1}{c|}{\color[HTML]{009901}\checkmark}      & \multicolumn{1}{c|}{0.549 (0.192)}                             & \multicolumn{1}{c|}{0.507 (0.184)}                      
\\ \cline{2-7} \cline{9-14} 

  & \color[HTML]{FE0000} --          & \color[HTML]{009901}\checkmark & \color[HTML]{FE0000} --          & \color[HTML]{FE0000} --          & 0.520 (0.104)                   & 0.502 (0.106)       
& &   &    &    &   &    &     \\ \hhline{|~|-|-|-|-|-|-|~|-|-|-|-|-|-|} 

  & \color[HTML]{FE0000} --          & \color[HTML]{FE0000} --          & \color[HTML]{009901}\checkmark & \color[HTML]{FE0000} --          & {0.484 (0.075)}          & \multicolumn{1}{c|}{{0.471 (0.076)}} &  & \multicolumn{2}{|c|}{\cellcolor{lightgray}}       & \multicolumn{2}{c|}{\cellcolor{lightgray}\textbf{GAZE}}   & \multicolumn{1}{c|}{\cellcolor{lightgray}\textbf{Accuracy}} & \multicolumn{1}{c|}{\cellcolor{lightgray}\textbf{F1}}  \\ \cline{2-7} \cline{11-14}

\multirow{-16}{*}{\textbf{EDA}} & \color[HTML]{FE0000} --          & \color[HTML]{FE0000} --          & \color[HTML]{FE0000} --          & \color[HTML]{009901}\checkmark & 0.425 (0.103)   & \multicolumn{1}{l|}{0.328 (0.116)} & & \multicolumn{2}{|c|}{\multirow{-2}{*}{\cellcolor{lightgray}\textbf{MOTION}}}      &   \multicolumn{2}{c|}{\color[HTML]{009901}\checkmark}       & \multicolumn{1}{l|}{0.529 (0.116)}     & \multicolumn{1}{c|}{0.484 (0.121)}     \\ \cline{1-7} \cline{9-14} 

\end{tabular}
}
\end{table}

To build the fusion models, we used the features extracted from 4-second windows with 2-second overlap.

As shown in Table~\ref{tab:fusion-results}, all fusion models surpassed the baseline performance of 17\%.
Surprisingly, the highest performance was not achieved by incorporating all sensor groups. While the comprehensive model including all sensors yielded an F1 score of 60.9\% (SD: 15.8\%) and accuracy of 63.7\% (SD: 14.3\%), excluding EDA produced better results with an F1 score of 64.5\% (SD: 16.1\%) and accuracy of 67.1\% (SD: 14.7\%). 
Combinations including GAZE with other sensor groups demonstrated consistently strong performance. The MOTION+PUPIL+GAZE configuration achieved the second-highest F1 score of 62.1\% (SD: 14.6\%), while adding EDA to this combination slightly reduced performance to an F1 score of 61.4\% (SD: 14.5\%). EEG+GAZE+PUPIL and EEG+GAZE+MOTION combinations yielded comparable results with F1 scores of 53.4\% (SD: 16.4\%) and 53.1\% (SD: 16.4\%), respectively.

The weakest performing combinations involved EDA paired with either EEG or GAZE, resulting in F1 scores of 23.4\% (SD: 9.9\%) and 32.8\% (SD: 11.6\%) with corresponding accuracies of 32.9\% (SD: 9\%) and 42.5\% (SD: 10.3\%), respectively. These combinations proved substantially less effective for the classification task.

Notably, the model using only physiological groups (EEG+EDA+PUPIL) achieved comparable performance to the model using only behavioral groups (GAZE+MOTION), with F1 scores of 49.9\% (SD: 14.7\%) and 48.4\% (SD: 12.1\%), and accuracies of 53.5\% (SD: 14.1\%) and 52.9\% (SD: 11.6\%), respectively. This suggests that physiological and behavioral groups contribute useful but distinct information to the classification task.

\section{Discussion \& Limitation}
This section broadly discusses our results and the implications and opportunities that it hints towards. It also identifies the limitations and discusses the ecological validity of the released dataset.

\subsection{RQ1. The Influence of Cognitive Intents on the Physiological Responses}

Our first research question investigates whether varying cognitive intents during different search stages influence the physiological responses.
To address this, we performed statistical analyses comparing all search stages (refer to Section~\ref{subsec:responses_stages}), and further analyses focusing on comparing \eyeopen, \ifn, and \rjr (refer to Section~\ref{subsec:influence_intents}) -- each representing eye-driven activities but with distinct intents. 

Our results showed that when comparing all stages, the EEG $Alpha$ band activity differs across search stages {at all regions, with particularly the highest statistical values observed at the} left-parietal (P7), right-frontal (F4) and left-occipital (O1) regions, reflecting shifts in attention due to the visual, audio or spatial processing demands, and in decision-making processing. The EEG $Beta$ also revealed differences across all regions, suggesting sustained cognitive engagement. The statistically significant differences in phasic EDA indicate arousal changes. 

Our findings might suggest that during stages of information searching, the participants were continuously engaged in active thinking and problem-solving cognitive processing to navigate and evaluate information. The statistically significant differences imply the influence of both interaction modalities and cognitive intents.

To examine this more closely, we then focused on the eye-driven stages: \eyeopen, \ifn, and \rjr, which have the same interaction modality but different intents. 
Our results that EEG $Alpha$ differed in temporal-parietal regions suggest that attentional focus and cognitive effort vary depending on the intent behind gaze movements.

Considering the possible cognitive activities participants might engage in during these stages, at \eyeopen, participants only need to focus on the screen and relax. 
At \ifn, participants may engage in cognitive processes such as comprehending information, retrieving memories, assessing personal interest, and updating knowledge gaps, which drive subsequent actions \cite{Dominika2022information, moshfeghi2018search}. 
At \rjr, beyond comprehension, participants also evaluated information based on factors such as usefulness and topical relevance, ultimately deciding whether to accept it \cite{allegretti2015relevance, ji2024characterizing, moshfeghi2018search}.
Taken together, and in alignment with existing findings, our results indicated the lack of significant difference between \ifn and \rjr implies these two tasks engage similar neural mechanisms, while \eyeopen appears to invoke distinct attentional shifts.

\subsection{RQ2. The Influence of Interaction Modalities on the Physiological Responses}

Our second research question explores whether different interaction modalities, at search stages, influence the physiological responses.
To address this, we performed pairwise comparisons between search stages that share the same cognitive intents but differ in interaction method, specifically, \qss and \qst, and \rjr and \rjl (refer to Section~\ref{subsec:influence_modalities}).

As mentioned earlier, our results showed that EEG $Alpha$ activity at the left-parietal (P7) and left-occipital (O1) regions exhibited the most prominent statistical values across all search stages, indicating dynamics in visual, auditory, or spatial processing demands.
This might suggest that when the intents are the same, the ways of interaction influence how the brain processes information, mainly related to parts that control the sensory interactions. 
The pairwise analysis revealed statistical differences in the EEG $Theta$ and $Alpha$ activities at the parietal-occipital regions (P7, O1, O2), and also the right-frontal regions (F4) between \rjr and \rjl. This might suggest that receiving the search results by reading or by listening primarily affects the cognitive processes related to memory, visuospatial or auditory processing, and temporal integration.

In this dataset, the lengths of the queries submitted by participants between \qst and \qss were similar. The text queries had an average of 8 words, ranging from 2 to 19 words, and the voice queries had an average of 7.8 words, ranging from 2 to 20 words. However, the interaction durations differed. Typing required more time (average: 17.7 seconds) compared to speaking (average: 6.8 seconds). These differences in interaction duration and underlying cognitive processing might manifest in changing physiological data \cite{iadarola2021analysis}.  
Interestingly, no significant differences were observed in any brain regions or peripheral physiological data between \qst and \qss. {This suggests that the modality used to express the query -- typed or spoken} -- may not significantly affect cognitive load or neural engagement captured by the physiological data.

For future research, depending on the research goals, we suggest that wearable EEG headsets should prioritize sensor placement at parietal-occipital sites to effectively track modality-related cognitive shifts. For real-time attention and workload estimation, adding frontal electrodes (e.g., F4) can help detect mental effort differences in information reception tasks, or integrating eye-tracking and EDA sensors in smart wearable devices. 
The wearable systems (e.g., AR glasses, brain-sensing headbands) \cite{Kosmyna2024watching} should adjust content delivery, optimizing input/output modalities for cognitive efficiency.

\subsection{RQ3. Classification Performance of Physiological and Behavioral Data}

Our third research question RQ3 examines the ability of physiological and behavioral data to discriminate between search stages. To address this, we develop machine-learning models and compare their performance in classifying search stages using both individual and combined sensor groups (refer to Section~\ref{sec:classification}).

Overall, the physiological data yielded varying classification performances. The model using only EDA data demonstrated low performance, with the average F1 scores only marginally better than the baseline. Then, the fusion models that incorporate EDA also showed reduced performance. These suggest that cognitive processes during search stages do not strongly influence wristband-captured EDA in this controlled setup.
EEG data demonstrated limited capacity in discriminating the search stages. The minimal impacts of feature window size might suggest that similar neural responses occur within short time frames and remain consistent. This aligns with EEG characteristic of capturing rapid neural events, where extending the time window provides little additional information. 
PUPIL data exhibited the best performance, with the single model and majority of the fusion models achieving average F1 scores exceeding 50\%. Pupillary response is usually used to indicate attention and cognitive load \cite{van2018pupil}. This high performance might imply varying inherent demands of cognitive load across search stages. Therefore, researchers should be cautious when using pupillary responses to investigate experimental factors in different contexts to ensure validity.

Behavioral data, including GAZE and MOTION, also discriminated certain search stages with moderate performance. But the skewed predictions shown in Figure~\ref{fig:cm-prediction-42} reflect that they mainly captured the variants in interaction modalities rather than cognitive intent. The fusion models combining these with physiological data led to minor improvements. This further suggests they are useful for context detection but limited for detecting subtle activities.

Besides, the improvements in GAZE, PUPIL, and MOTION models with longer feature windows suggest these modalities benefit from capturing complete behavioral sequences. Ocular and motor responses likely unfold over longer periods and contain meaningful temporal patterns that shorter windows fail to detect. For instance, complete gaze patterns, pupillary reactions to cognitive load, and motion sequences associated with different tasks may require several seconds to fully manifest their distinctive characteristics.

Overall, the differential performance across sensor groups further validates their varying effectiveness in this classification task, and underscores the importance of sensor selection based on the specific cognitive processes under investigation.
And the impacts of feature window sizes suggest that optimal feature extraction windows should be tailored to each sensor's specific temporal dynamics rather than applying uniform parameters across all modalities. For real-time applications, this may necessitate modality-specific processing pipelines with different temporal configurations to maximize classification performance.

\subsection{Limitations \& Ecological Validity}
\label{subsec:limit_validaity}

Our dataset serves as a baseline for understanding and assessing consumer-grade sensor data during information searching in a controlled lab setting. However, we note a number of limitations that researchers should be aware of when using the dataset.
The size of the \dataset is relatively small compared to other sensor datasets, with only 20 participants, where young adults are over-represented. While tasks were presented in English and all participants had professional working proficiency in English, the majority of participants are not native speakers. 
The dataset only contains eye-tracking data from 18 participants due to a device error. In addition, tasks like speaking, listening, and typing did not necessarily require participants to focus on the screen. This naturally resulted in missing eye-tracking data, which reflects inherent limitations of applying eye-tracking to such tasks.

The data was collected in a controlled lab environment with simulated searches, which may not fully represent real-world distractions and variability. 
In a natural setting, sensing data can be affected by factors such as contents, visual elements, subjective perceptions, and external environments.
We focused specifically on information searching scenarios while controlling variables such as the type of information need (search goal), topic difficulty, familiarity, information complexity, and relevance using a simplified interaction system. While this approach may not perfectly reflect real-life user behaviors, it provides clearer insights into how sensing data responds during information activities. This methodology helps analyze distinct search stages, though it might not capture how these stages seamlessly integrate in everyday use.

Despite our controls, some intricate activities retained variability in their natural state, revealing the dynamic nature of human-information interaction. For example, we separated query formulation from submission to isolate underlying processes. This acknowledges differences between voice interactions (where users may pause before speaking) and text-based interactions (where thinking and typing occur simultaneously). In our study, some participants reported mind-wandering during query formulation, while others used the time for focused thinking, and a few needed to recall their information need before forming queries. As a result, cognitive processes and corresponding physiological responses still overlapped across search stages.

Moreover, search stage durations varied considerably, from 6.8 to 66.0 seconds on average, highlighting the dynamic nature of information searching and the importance of context-specific factors. This variability likely affected physiological responses, as longer stages may elicit more complex responses than shorter ones.
In contrast to most existing physiological datasets, which typically involve longer engagement periods, this duration variability could benefit studies featuring rapid cognitive transitions of short-term interactions that characterize real-world search interactions.

Lastly, our study employed only consumer-grade sensors that offer lower accuracy and sampling rates than medical-grade equipment; hence, there is a possibility of artifacts or missed responses that may have impacted the results. As a trade-off, these sensors are closer to the actual technologies users encounter in everyday settings.
Future research could bridge this gap by implementing a dual-monitoring approach, using both consumer- and medical-grade or more advanced equipment simultaneously to validate findings while maintaining ecological validity. Additionally, larger sample sizes would strengthen statistical analysis and help distinguish true physiological patterns.
Advantages of consumer-grade sensors, such as accessibility, comfort, and ease of use, allow for a scalable version of this dataset.

A future extension of the \dataset dataset could focus on featuring participants 1) of an older age, 2) with their native language, and 3) with neuro-diversity features, or featuring the tasks with 4) complex information needs, 5) personalized search results (e.g., conversational interaction), and 6) multi-turn interactions.

\section{Conclusion and Future Work}

We presented SenseSeek, a comprehensive dataset examining the ramifications of passive sensing data and information activity,  particularly focusing on the information search process. 
The dataset comprises physiological and behavioral data from 20 participants across 235 trials, 940 search stages in a controlled lab environment, where each participant completed simulated search tasks. Both the raw and cleaned data are made publicly available, along with 258 extracted features. To provide contextual richness, it also contains the task materials, self-rated perceptions, and gaze-annotated screen recordings.
We validate the usefulness of the dataset by investigating how cognitive intents and interaction modalities affect information searching stages and {evaluating} the effectiveness of sensor data in distinguishing these stages. Statistically significant differences were observed in EEG and PUPIL data, particularly influenced by cognitive intents and interaction modality (reading vs. listening).
Search stages classification was performed using features from individual or multiple sensor groups. Our analysis showed that PUPIL features performed best, and EDA features performed worst. 
The varying effectiveness of sensor groups underscores the need for sensor selection based on specific cognitive processes and modality-specific processing configurations to optimize classification performance in multi-modality models.

Future work can further explore the relationship between the physiological signals and the variables involved in search stages. In particular, the SenseSeek dataset can be used to advance knowledge on characterizing the perception of topic familiarity, topic difficulty, and relevance with multiple physiological signals.

By making this dataset publicly available and {deepening} our understanding of these variations, we hope to contribute to the development of passive sensors and multimedia information access systems. For instance, designing adaptive search interfaces that respond to users' cognitive needs in real-time.
This dataset lays a foundation for future research exploring more complex search activities and provides a rich source of data for developing algorithms that monitor user experiences such as attention, workload, and stress during information interaction. Researchers can leverage this baseline data to train computational models before applying them in naturalistic environments, ensuring more reliable transitions from controlled to real-world applications.

\section{Dataset and Code Availability}
The use of the \dataset dataset is limited to academic research purposes. The \dataset dataset is publicly available and can be accessed on the OSF platform at 
\url{https://osf.io/waunb/?view_only=94756f9d2c7a49e094ae42d494c9516a}.
{Additional descriptions of the dataset are included in the appendices.}
The code used for data processing, analysis, and model training is 
also available at 
\url{https://github.com/ADMSCentre/SenseSeek-Dataset-code}.
The repository contains the instructions for the dataset, several Jupyter Notebook files with data analysis and visualizations, and Python source files for model training. All required packages are listed in the \texttt{requirements.txt} file.

\begin{acks}
{This research is partially supported by the \grantsponsor{ARC}{Australian Research Council}{https://www.arc.gov.au/} and the \grantsponsor{ADM+S}{Australian Research Council Centre of Excellence for Automated Decision-Making and Society}{https://www.admscentre.org.au/} (\grantnum{ARC}{DE200100064}, \grantnum{ARC}{CE200100005}).}
\end{acks}

\newpage

\balance
\bibliographystyle{ACM-Reference-Format}
\bibliography{reference}

\newpage
\appendix

\section{Additional Description of the \dataset dataset}
\label{apx:add_description}

\begin{figure}[htbp!]
    \centering
    \begin{subfigure}[t]{0.33\linewidth}
        \fbox{\includegraphics[width=0.95\linewidth]{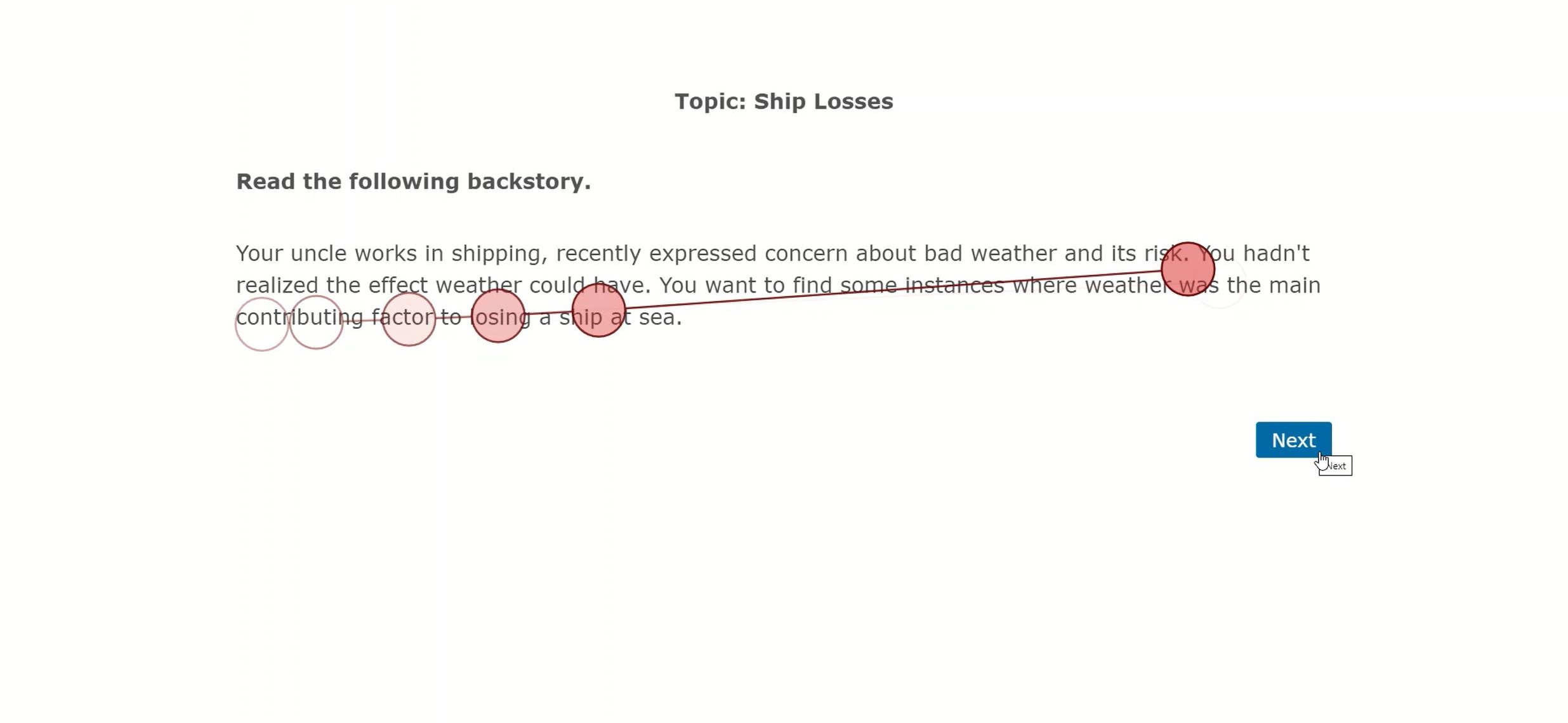}}
    \end{subfigure}
    ~
    \begin{subfigure}[t]{0.33\linewidth}
        \fbox{\includegraphics[width=0.95\linewidth]{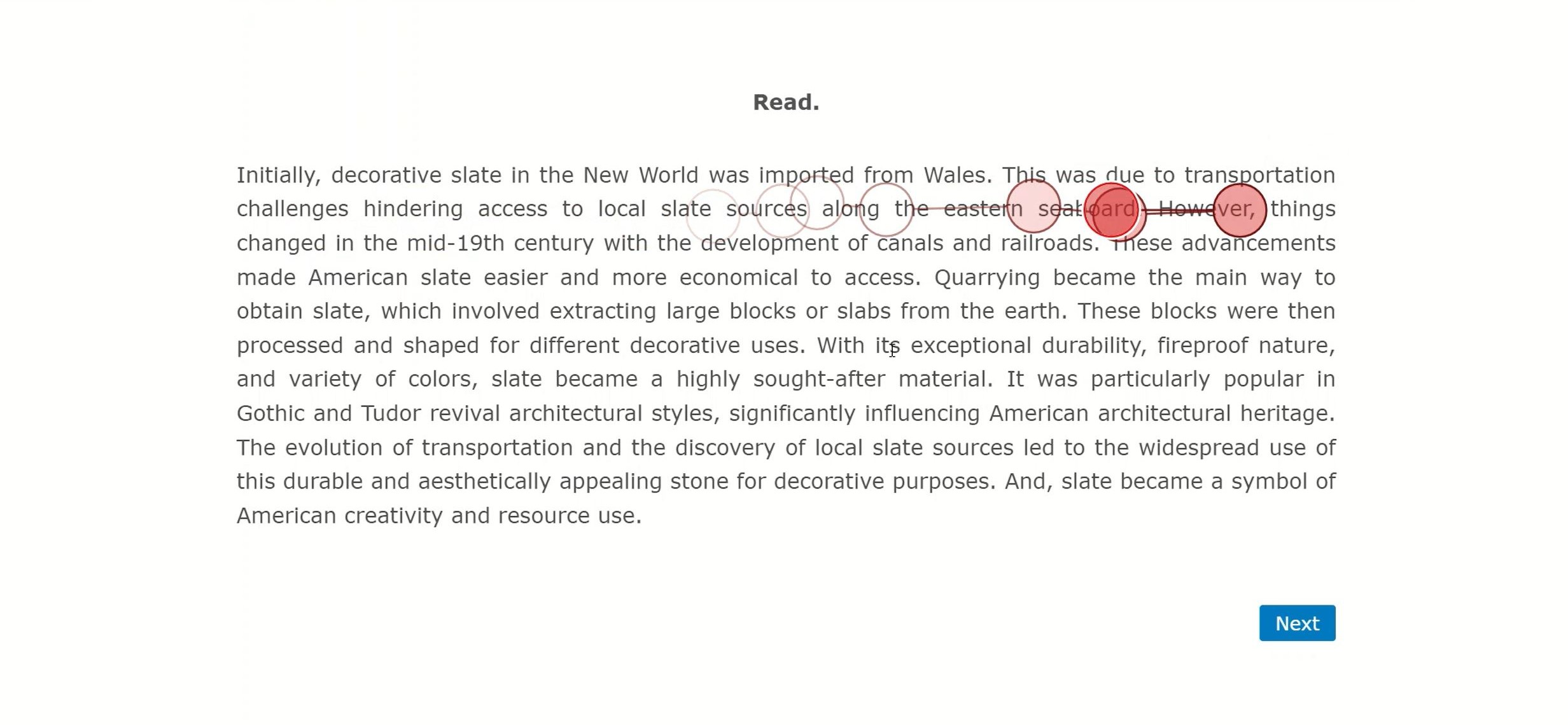}}
    \end{subfigure}
    ~
    \begin{subfigure}[t]{0.33\linewidth}
        \fbox{\includegraphics[width=0.95\linewidth]{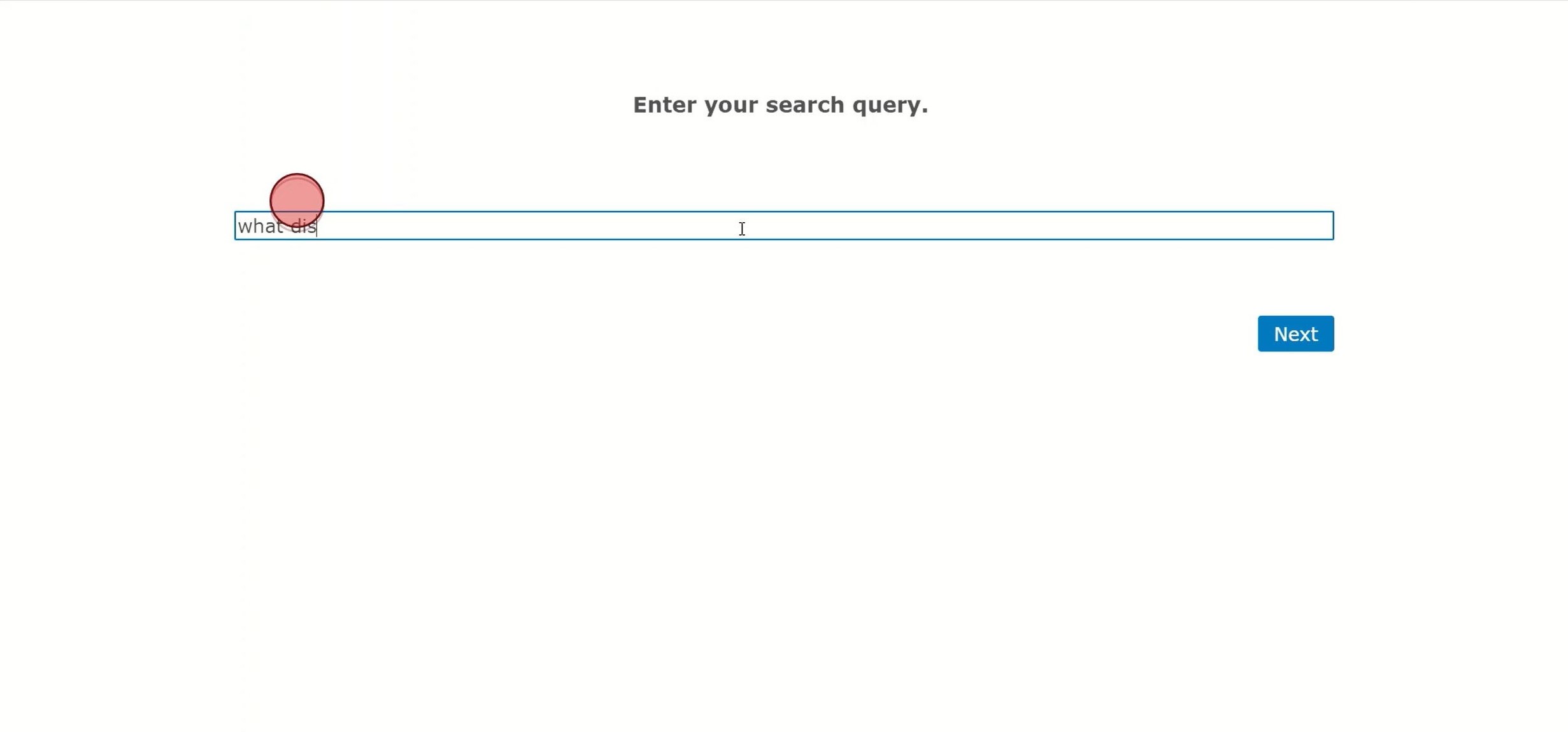}}
    \end{subfigure}
    \caption{Examples of Gaze-annotated screen recordings.}
    \Description{Screenshots from the experiment showing gaze tracking overlays on the interface. These visuals demonstrate how participants visually engaged with the content.}
\end{figure}

\begin{table}[ht]
\caption{Demographic information \& pre-study questionnaire per participant.}
\resizebox{\linewidth}{!}{
\begin{tabular}{cccccccc}
\toprule
\textbf{PID} & \textbf{\begin{tabular}[c]{@{}c@{}}Age \\ (years old)\end{tabular}} & \textbf{Gender} & \textbf{English Proficiency Level} & \textbf{Right-handed} & \textbf{\begin{tabular}[c]{@{}c@{}}Sleep \\ (hours)\end{tabular}} & \textbf{\begin{tabular}[c]{@{}c@{}}Caffeine Intake\\ (cups, in the last two hours)\end{tabular}} & \textbf{Wear Glasses} \\ \midrule
PA5  & 18--24 & Female & Native English & Y & 7--9 & N & N \\
PA6  & 25--34 & Male & Native English & Y & \textless 7 & Y--1 & N \\
PA8  & 18--24 & Female & Professional Work English & Y & \textless 7 & N & Y \\
PA9  & 25--34 & Female & Professional Work English & N & 7--9 & Y -- 1 & N \\
PA11 & 25--34 & Male & Professional Work English & Y & \textless 7 & N & Y \\
PA12 & 25--34 & Male & Professional Work English & N & 7--9 & N & N \\
PA13 & 18--24 & Male & Full Professional English & Y & \textless 7 & Y -- 1 & N \\
PA17 & 18--24 & Male & Professional Work English & Y & 7--9 & N & Y \\
PA18 & 25--34 & Female & Professional Work English & Y & 7--9 & Y $\geq$ 2 & Y \\
PA19 & 25--34 & Male & Professional Work English & Y & 7--9 & N & N \\
PA20 & 25--34 & Male & Professional Work English & Y & \textless 7 & N & N \\
PA21 & 25--34 & Female & Professional Work English & Y & 7--9 & N & Y \\
PA22 & 25--34 & Male & Professional Work English & Y & 7--9 & N & Y \\
PA27 & 25--34 & Male & Full Professional English & Y & 7--9 & N & Y \\
PA29 & 35--44 & Male & Professional Work English & Y & 7--9 & N & N \\
PA30 & 25--34 & Male & Full Professional English & Y & 7--9 & N & Y \\
PA31 & 25--34 & Female & Native English & Y & 7--9 & N & N \\
PA32 & 25--34 & Male & Native English & Y & 7--9 & N & N \\
PA33 & 25--34 & Female & Native English & Y & 7--9 & N & N \\
\bottomrule
\end{tabular}
}
\end{table}

\subsection{Details of Feature Extraction}
\label{apx_sub:features}

\paragraph*{EDA}
A total of 31 EDA features were extracted, following \citet{ngage2020gao}, \citet{barral2015exploring}, and \citet{bias2023boon}. 
We extracted the 5 descriptive statistical features, including mean, minimum, maximum, standard deviation, and range, from the mixed EDA values and their first and second derivatives, the Skin Conductance Level (SCL) and the Skin Conductance Response (SCR). Apart from that, we also extracted more statistical features, including median, variance, skewness, and kurtosis, from the mixed EDA values, median, and mean amplitude for the SCL and SCR.

\paragraph*{EEG}
A total of 180 EEG features were extracted, following \citet{Gwizdka2017temporal} and \citet{bird2018study}. We used the \textit{eeglib} library~\cite{CABANEROGOMEZ2021100745} for feature extraction.

For each of the 14 channels, we computed the mean, standard deviation, kurtosis, curve length, zero crossings, number of peaks, and wavelet entropy, which are the common features in most EEG experiments \cite{Gwizdka2017temporal, koelstra2011deap}. We also calculated the powers of the 4 frequency bands for each of the 14 EEG channels \cite{bird2018study}, $Theta$ (4--8 Hz), $Alpha$ (8--13 Hz), $Beta$(13--25 Hz) and $LowGamma$ (25--40 Hz) bands. The powers were normalized by dividing by the total powers. 

Additionally, we derived features for left-right hemispheric asymmetry in each frequency band. Following \citet{Gwizdka2017temporal}, we divided the channels into four groups: frontal-left (AF3, F7, F3, FC5), frontal-right (FC6, F4, F8, AF4), back-left (P7, O1), and back-right (P8, O2). Inter-hemispheric asymmetry was calculated by dividing the mean power of right hemisphere groups (frontal-right and back-right) by the mean power of left hemisphere groups (frontal-left and back-left). The intra-hemispheric asymmetry was calculated by dividing the mean power of the frontal group by the mean power of the back group for each hemisphere.

\paragraph*{PUPIL}
A total of 25 PUPIL features extracted, following \citet{Gwizdka2017temporal}.
We extracted the 5 descriptive statistical features, including mean, minimum, maximum, standard deviation and range from pupil diameter, Relative Pupil Dilation (RPD), and the first derivatives of RPD \cite{Gwizdka2017temporal}. We also extracted the 5 quartiles ($10th$, $25th$, $50th$, $75th$, $90th$) from the RPD and the 3 quartiles ($25th$, $50th$, $75th$) from the first derivatives of RPD \cite{Gwizdka2017temporal}.
Apart from that, the LHIPA index was also extracted. It is a frequency-based approach to measure cognitive load from pupil diameter, purposed by ~\citet{Duchowski2020lowhigh}.  

\paragraph*{MOTION}
A total of 13 MOTION features were extracted from the wrist motion and head motion. 
The 5 descriptive statistical features of wrist motion were extracted using magnitude~\cite{ngage2020gao}. 
For head motion features, we extracted the mean of the $x,y,z$ coordinates on both angular speed and acceleration. Besides, we also extracted the mean and energy from the magnitude of the acceleration.

\paragraph*{GAZE}
A total of 6 GAZE features were extracted, following \citet{Gwizdka2017temporal}.
The GAZE features include the number of occurrences, total and mean duration of each type of eye movement, i.e., fixation and saccade.

\newpage
\section{Additional Materials used in Data Collection}
\label{apx:add_material}

\subsection{Preparation of the Topics \& Backstories \& Search Results}
The topics and the corresponding backstories (i.e., task scenarios) are selected from the TREC2002-InformationNeed dataset~\cite{informationneed2014}. This dataset contains backstories corresponding to different topics from the TREC2002-4 topic set, and categorizes them into three labels in terms of cognitive complexity.
We use the topics from the Understanding category, which requires the participants to find information and gain some understanding of the topics. 
After removing the topics related to crises, wars, conspiracy, or politics, which might trigger subjective feelings, we select 12 topics.

We use GPT-3.5 to slightly edit the backstories to ensure they all have similar word counts. As a result, the backstories have an average of 40$\pm$1 words.

For each backstory, we manually select 2 -- 3 articles from TREC document collections, which are relevant to the information need. GPT-3.0 is used to generate the search result based on the provided articles and a binary factual judgment question. The judgment question is to ensure the participants have engaged in the task. We further manually examine the search results and check with the Flesch-Kincaid Grade readability level, and edit with GPT-3.5 if it is too easy or too hard.
All of them are converted into an audio format with the Google text-to-speech API for the listening tasks.

\begin{table}[htbp!]
\caption{List of Topics and Backstories used. For the search results and more information, please find them in the dataset file. *Text REtrieval Conference (TREC) Data: \url{https://pages.nist.gov/trec-browser/}}
\resizebox{\linewidth}{!}{
\begin{tabular}{@{}clllcccc@{}}
\toprule
\textbf{\begin{tabular}[c]{@{}c@{}}Topic\\ ID\end{tabular}} & \multicolumn{1}{c}{\textbf{\begin{tabular}[c]{@{}c@{}}TREC*\\ ID\end{tabular}}} & \multicolumn{1}{c}{\textbf{Topic}}                                         & \multicolumn{1}{c}{\textbf{Backstory (revised with GPT3.5)}}                                                                                                                                                                                                                                                    & \textbf{\begin{tabular}[c]{@{}c@{}}Backstory \\ word count\end{tabular}} & \textbf{\begin{tabular}[c]{@{}c@{}}Backstory ID \\ (from \cite{informationneed2014}))\end{tabular}} & \textbf{\begin{tabular}[c]{@{}c@{}}Search Result \\ word count\end{tabular}} & \textbf{\begin{tabular}[c]{@{}c@{}}Search Result\\ (Flesch-Kincaid \\ Readability)\end{tabular}} \\ \midrule
314                                                         & R03.314                                                                        & \begin{tabular}[c]{@{}l@{}}Marine \\ Vegetation\end{tabular}               & \begin{tabular}[c]{@{}l@{}}You recently heard a commercial about the health benefits of eating algae, seaweed and kelp. \\ This made you interested in finding out about the positive uses of marine vegetation, both as \\ a source of food, and as a potentially useful drug.\end{tabular}                    & 41                                                                       & IN14.002                                                                                            & 143                                                                          & 12.6                                                                                             \\
320                                                         & R03.320                                                                        & \begin{tabular}[c]{@{}l@{}}Undersea Fiber\\ Optic Cable\end{tabular}       & \begin{tabular}[c]{@{}l@{}}The FLAG (Fiber optic link around the globe) system will be the world's longest undersea \\ fiber optic cable. You are interested in finding out more about the project, including which \\ companies are involved, and what technology is needed for such an endeavor.\end{tabular} & 42                                                                       & IN14.164                                                                                            & 146                                                                          & 10.5                                                                                             \\
353                                                         & R03.353                                                                        & \begin{tabular}[c]{@{}l@{}}Antarctica \\ Exploration\end{tabular}          & \begin{tabular}[c]{@{}l@{}}On the T.V. news last night, you saw footage of scientists in Antarctica. There seemed a \\ surprising number of people there. This got you wondering what scientific expeditions or \\ projects are under way in Antarctica, and what is planned.\end{tabular}                    & 39                                                                       & IN14.018                                                                                            & 147                                                                          & 10.8                                                                                             \\
355                                                         & R03.355                                                                        & \begin{tabular}[c]{@{}l@{}}Ocean Remote\\ Sensing\end{tabular}             & \begin{tabular}[c]{@{}l@{}}A friend at a university is excited about a chance to work with satellite data for ocean \\ remote sensing. You've become interested in this and you'd like to find out what's \\ being developed in this field and how it's being used.\end{tabular}                               & 41                                                                       & IN14.102                                                                                            & 148                                                                          & 11.5                                                                                             \\
416                                                         & R03.416                                                                        & \begin{tabular}[c]{@{}l@{}}Three Gorges\\ Project\end{tabular}             & \begin{tabular}[c]{@{}l@{}}A friend is traveling to China and plans to cruise the Yangtze River.  You're not sure whether\\  that's still possible, as it might now be dammed. You'd like to find out the current status of \\ the Three Gorges Dam project.\end{tabular}                                       & 40                                                                       & IN14.077                                                                                            & 145                                                                          & 12.8                                                                                             \\
419                                                         & R03.419                                                                        & \begin{tabular}[c]{@{}l@{}}Recycle, \\ Automobile Tires\end{tabular}       & \begin{tabular}[c]{@{}l@{}}You need to buy new tires for your car, and the local dealer has offered to take the old ones \\ for recycling. You didn't know tires could be recycled and you wonder what new uses they \\ are being put to.\end{tabular}                                                          & 40                                                                       & IN14.084                                                                                            & 147                                                                          & 12.6                                                                                             \\
433                                                         & R03.433                                                                        & \begin{tabular}[c]{@{}l@{}}Greek, Philosophy, \\ Stoicism\end{tabular}     & \begin{tabular}[c]{@{}l@{}}You visited a museum recently, and heard about the Greek philosophy of stoicism. You \\ start wondering if there is any contemporary interest in this philosophy, and whether \\ Greek stoic plays and artistic productions are taking place or being planned.\end{tabular}        & 39                                                                       & IN14.146                                                                                            & 149                                                                          & 11                                                                                               \\
448                                                         & R03.448                                                                        & Ship Losses                                                                & \begin{tabular}[c]{@{}l@{}}Your uncle works in shipping, recently expressed concern about bad weather and its risk. \\ You hadn't realized the effect weather could have. You want to find some instances where\\  weather was the main contributing factor to losing a ship at sea.\end{tabular}               & 41                                                                       & IN14.035                                                                                            & 152                                                                          & 12.3                                                                                             \\
708                                                         & T04.708                                                                        & \begin{tabular}[c]{@{}l@{}}Decorative \\ Slate Sources\end{tabular}        & \begin{tabular}[c]{@{}l@{}}Your sister has just moved in to a new house, with slate stone flooring and a slate stone\\ counter top. You are now curious to explore the sources of slate stone for decorative use,\\ and how it is obtained.\end{tabular}                                                        & 39                                                                       & IN14.033                                                                                            & 150                                                                          & 12.6                                                                                             \\
743                                                         & T04.743                                                                        & \begin{tabular}[c]{@{}l@{}}Freighter Ship\\  Registration\end{tabular}     & \begin{tabular}[c]{@{}l@{}}You are thinking of going into the shipping business, and decide that you want to learn \\ more about the regulations framework. Specifically, you aim to explore rules and \\ considerations related to the registration of a freight ship in a country.\end{tabular}               & 40                                                                       & IN14.110                                                                                            & 146                                                                          & 13.4                                                                                             \\
711                                                         & T04.711                                                                        & \begin{tabular}[c]{@{}l@{}}Train Station \\ Security Measures\end{tabular} & \begin{tabular}[c]{@{}l@{}}There has been a great deal of publicity given to security at airports. But many people also \\ travel by train. You are interested in understanding what security measures have been\\employed at train stations due to increased security concerns.\end{tabular}                 & 39                                                                       & IN14.142                                                                                            & 146                                                                          & 11.8                                                                                             \\
725                                                         & T04.725                                                                        & \begin{tabular}[c]{@{}l@{}}Low White \\ Blood Cell Count\end{tabular}      & \begin{tabular}[c]{@{}l@{}}You've recently been to your doctor and learned you have a low white blood cell count, but\\  it's not clear yet why this is. You want to find out what disease or condition might have \\ caused this symptom.\end{tabular}                                                         & 38                                                                       & IN14.059                                                                                            & 155                                                                          & 11.3                                                                                             \\ \bottomrule
\end{tabular}}
\end{table}

\begin{table}[htbp!]
\caption{List of self-rating items. }
\resizebox{\linewidth}{!}{
\begin{tabular}{@{}clll@{}}
\toprule
                                                                                                  & \multicolumn{1}{c}{\textbf{Self-rating Item}} & \multicolumn{1}{c}{\textbf{Text}}                    & \multicolumn{1}{c}{\textbf{Values}}                                                                                                                                       \\ \midrule
\multirow{3}{*}{\textbf{\begin{tabular}[c]{@{}c@{}}Set 1\\ (on the Topic)\end{tabular}}}          & \textit{Topic Difficulty}                     & I am familiar with this topic.                       & \multirow{5}{*}{\begin{tabular}[c]{@{}l@{}}1: Strongly disagree\\ 2: Somewhat disagree\\ 3: Neither agree nor disagree\\ 4: Somewhat agree\\5: Strongly agree\end{tabular}} \\
                                                                                                  & \textit{Topic Interest}                       & I am interested in this topic.                       &                                                                                                                                                                           \\
                                                                                                  & \textit{Topic Familiarity}                    & I feel difficult to understand this topic.           &                                                                                                                                                                           \\ \cmidrule(r){1-3}
\multirow{2}{*}{\textbf{\begin{tabular}[c]{@{}c@{}}Set 2 \\ (on the Search Result)\end{tabular}}} & \textit{Information Relevance}                & This information is relevant to my information need. &                                                                                                                                                                           \\
                                                                                                  & \textit{Information Difficulty}               & I feel difficult to understand this information.     &                                                                                                                                                                           \\ \bottomrule
\end{tabular}
}
\end{table}

\newpage
\section{Self-Rated Task Perceptions per Topic/Participant}

\begin{table}[ht]
\caption{Average scores of self-ratings for 12 topics.}
\resizebox{\linewidth}{!}{
\begin{tabular}
{p{.079\linewidth}|p{.077\linewidth}|p{.077\linewidth}|p{.077\linewidth}|p{0.07\linewidth}|p{0.06\linewidth}|p{.077\linewidth}|p{0.093\linewidth}|p{0.055\linewidth}|p{0.079\linewidth}|p{0.08\linewidth}|p{0.089\linewidth}|p{0.08\linewidth}}
\toprule
\textbf{ID}           & \textbf{314}               & \textbf{320}                        & \textbf{353}                    & \textbf{355}                  & \textbf{416}                  & \textbf{419}                      & \textbf{433}                       & \textbf{448}         & \textbf{708}                      & \textbf{743}                         & \textbf{711}                             & \textbf{725}                        \\ \midrule
\textbf{Topic}              & \textbf{\small Marine Vegetation} & \textbf{\small Undersea Fiber Optic Cable} & \textbf{\small Antarctica exploration} & \textbf{\small Ocean Remote Sensing} & \textbf{\small Three Gorges Project} & \textbf{\small Recycle Automobile Tires} & \textbf{\small Greek Philosophy Stoicism} & \textbf{\small Ship Losses} & \textbf{\small Decorative Slate Sources} & \textbf{\small Freighter ship registration} & \textbf{\small Train Station Security Measures} & \textbf{\small Low White Blood Cell Count} \\ \midrule
\parbox{\linewidth}{\small \textbf{topic\_\\difficulty}}   & 2.2   & 2.4   & 2.4   & 2.4   & 2.7   & 2.2   & 2.8   & 2.3   & 2.7   & 2.2   & 2.6   & 2.8   \\ \midrule
\parbox{\linewidth}{\small \textbf{topic\_\\familiarity}} & 2.7   & 3.6   & 3.1   & 3.0   & 1.7   & 3.1   & 2.8   & 2.7   & 1.5   & 3.2   & 3.0   & 2.0   \\ \midrule
\parbox{\linewidth}{\small \textbf{topic\_\\interest}}    & 3.4   & 4.0   & 3.7   & 3.8   & 3.0   & 3.6   & 3.6   & 3.4   & 2.4   & 3.8   & 3.9   & 3.0   \\ \midrule
\parbox{\linewidth}{\small \textbf{info\_\\difficulty}}   & 1.8   & 2.2   & 1.7   & 2.0   & 1.5   & 2.0   & 2.3   & 1.7   & 2.0   & 1.4   & 2.2   & 2.5   \\ \midrule
\parbox{\linewidth}{\small \textbf{info\_\\relevance}}    & 3.4   & 4.2   & 4.2   & 4.2   & 3.7   & 4.3   & 3.4   & 4.2   & 3.8   & 4.6   & 4.4   & 3.2  \\ \midrule
\parbox{\linewidth}{\textbf{\small attention\_\\check}}    & 100\%   & 75\%& 100\%    & 95\%    & 90\%       & 85\%    & 55\%     & 95\%   & 75\%     & 75\%      & 70\%    & 55\%     \\
\bottomrule
\end{tabular}}
\end{table}

\begin{table}[ht]
\caption{Average scores of self-ratings for each participant.}
\resizebox{\linewidth}{!}{
\begin{tabular}{@{}lcccccc@{}}
\toprule
\textbf{PID}  & \textbf{topic\_familiarity} & \textbf{topic\_interest} & \textbf{topic\_difficulty} & \textbf{info\_revelance} & \textbf{info\_difficulty} & \textbf{attention\_check} \\ \midrule
\textbf{PA5}  & 1.8                         & 3.7                      & 4.0                        & 4.6                      & 1.2                       & 83\%                      \\
\textbf{PA6}  & 2.8                         & 3.8                      & 1.2                        & 4.1                      & 1.2                       & 100\%                     \\
\textbf{PA8}  & 3.1                         & 3.8                      & 2.5                        & 4.4                      & 1.6                       & 67\%                      \\
\textbf{PA9}  & 3.2                         & 3.2                      & 2.0                        & 4.6                      & 1.5                       & 83\%                      \\
\textbf{PA11} & 1.7                         & 3.2                      & 3.3                        & 4.5                      & 1.7                       & 67\%                      \\
\textbf{PA12} & 3.6                         & 3.4                      & 2.4                        & 3.4                      & 2.4                       & 75\%                      \\
\textbf{PA13} & 2.3                         & 3.9                      & 2.5                        & 3.6                      & 1.9                       & 83\%                      \\
\textbf{PA17} & 2.6                         & 3.6                      & 2.8                        & 3.8                      & 3.3                       & 92\%                      \\
\textbf{PA18} & 2.9                         & 3.6                      & 2.0                        & 4.2                      & 2.2                       & 25\%                      \\
\textbf{PA19} & 3.4                         & 4.0                      & 1.8                        & 4.3                      & 1.5                       & 83\%                      \\
\textbf{PA20} & 2.9                         & 3.2                      & 3.0                        & 3.8                      & 2.4                       & 67\%                      \\
\textbf{PA21} & 1.3                         & 1.7                      & 3.2                        & 3.7                      & 2.4                       & 67\%                      \\
\textbf{PA22} & 3.3                         & 3.9                      & 2.1                        & 3.3                      & 1.2                       & 83\%                      \\
\textbf{PA26} & 1.7                         & 1.8                      & 4.2                        & 3.9                      & 3.2                       & 83\%                      \\
\textbf{PA27} & 2.9                         & 3.9                      & 2.3                        & 3.7                      & 1.9                       & 92\%                      \\
\textbf{PA29} & 2.8                         & 4.0                      & 2.2                        & 3.8                      & 2.1                       & 100\%                     \\
\textbf{PA30} & 3.2                         & 4.0                      & 1.0                        & 3.6                      & 1.1                       & 100\%                     \\
\textbf{PA31} & 2.2                         & 3.3                      & 2.4                        & 4.5                      & 1.8                       & 92\%                      \\
\textbf{PA32} & 2.8                         & 3.8                      & 2.9                        & 3.6                      & 2.8                       & 75\%                      \\
\textbf{PA33} & 2.9                         & 3.1                      & 2.2                        & 4.2                      & 1.7                       & 100\%                     \\ \bottomrule
\end{tabular}
}
\end{table}

\end{document}
\endinput